\documentclass[10pt]{article}
\usepackage{graphicx}
  
\usepackage[table]{xcolor} 
\usepackage{bbm}
\usepackage{graphicx}
\usepackage[linesnumbered,ruled]{algorithm2e}
\usepackage{algpseudocode}
\usepackage{appendix}
\usepackage[utf8]{inputenc}
\usepackage{amsmath}
\usepackage{amsfonts}
\usepackage{amssymb}
\usepackage{mathtools}  % Formulas perks
\usepackage[version=4]{mhchem}
\usepackage{stmaryrd}
\usepackage{bbold}
\usepackage{enumitem}
\setlist[itemize]{label=\textbullet}
\usepackage{subcaption}
\usepackage{authblk}    % Authors and affiliations
\usepackage[margin=0.8in]{geometry}
\usepackage[numbers]{natbib}
\usepackage{hyperref}
\usepackage{multirow}
\usepackage{longtable}
\usepackage{booktabs}
\usepackage{tabularx}
\usepackage{array}
\usepackage{siunitx}

\begin{document}
\title{Simulated Annealing for Model-Robust Partial Profile Choice Designs in Healthcare Preference Studies}
 \author[1,2,*]{Yicheng Mao}
  \author[2,3]{Roselinde Kessels}
  \affil[1]{Department of Data Analytics and Digitalization, Maastricht University, P.O. Box 616, 6200 MD Maastricht, The Netherlands}
   \affil[2]{Department of Mathematics and Statistics, University of Calgary,
University Drive NW, Calgary, T2N 1N4, Canada}
  \affil[3]{Department of Economics, City Campus, University of Antwerp, Prinsstraat 13, 2000 Antwerp, Belgium}
  \affil[*]{Correspondence:
yicheng.mao1@ucalgary.ca}
\maketitle
\begin{abstract}
Discrete Choice Experiments (DCEs) investigate participants' preferences by observing their choice behavior in hypothetical scenarios and are widely used in the domain of healthcare. To reduce participants' cognitive burden, especially when dealing with a large number of attributes, researchers often employ partial profile designs. In these designs, certain attributes within each choice set are kept constant. Current literature on partial profile designs mainly focuses on main-effects models rather than interaction-effect models, with certain partial profile designs even incapable of estimating interaction effects. To address this issue, this paper introduces an Simulated Annealing (SA) algorithm to construct partial profile designs based on an interaction-effects model. During the experimental design phase, the existence and magnitude of interaction effects are often unknown. Therefore, this paper proposes a model-robust experimental design strategy. Through extensive simulation experiments and a real-life case study, we demonstrate that our SA model-robust partial profile design performs relatively well regardless of the underlying model.
\noindent 
\\ 

\end{abstract}
\section{Introduction}\label{Sec:Introduction}
Discrete choice experiments (DCEs) investigate individual preferences for attributes of products or services and have been widely applied in healthcare \citep{clark2014discrete,Jaynes2016Using}. In these experiments, participants are typically asked to select their preferred profile from a choice set composed of two or more profiles. These profiles are defined as combinations of levels of attributes. By analyzing the choice behavior of participants, researchers can gain insights into people's preferences for different attributes and levels, and make predictions about their decisions in real-life scenarios.

Many DCEs in healthcare study a large number of attributes, including examples such as \cite{sculpher2004patients}, \cite{Hall2006} and \cite{witt2009designing}. However, this abundance of attributes can result in an overload of information for respondents, significantly increasing their cognitive burden \citep{green1978conjoint}. To simplify the choice task, \cite{chrzan2010using} suggested that in each choice set, respondents should be provided with profiles varying in only a limited set of attributes, while keeping the levels of the other attributes constant. These types of profiles are known as partial profiles, as opposed to full profiles, where all the attributes are varying in each choice set. 
Typically, the number of varying attributes in partial profiles is consistent across each choice set and is known as the profile strength \citep{grasshoff2003optimal}. 
Employing a partial profile design can effectively reduce response error arising from respondents' inattention, thereby achieving improved response efficiency compared to a full profile design \citep{johnson2013constructing,Jonker2018,meyerhoff2023performance}. For example, \cite{Jonker2019} found in a randomized controlled DCE that using a partial profile design, compared to a full profile design, reduced the dropout rate by 30\% and increased the level of choice consistency by 33\%.

Due to the nonlinear nature of the choice model, the construction of optimal partial profile designs for DCEs necessitates knowledge of the model parameters' values.
However, these parameter values are often unknown at the experimental design stage.
To address this, three approaches have been proposed. The first approach, known as the utility-neutral design, assumes that all unknown parameter values are zero. While straightforward, this method relies on an unrealistic assumption that respondents exhibit no distinct preference for different attribute levels.
The second approach, locally optimal design, assigns predetermined nonzero values to the parameters \citep{huber1996importance}. Although more refined than the utility-neutral design, this method remains sensitive to the accuracy of the specified parameter values.
The third approach, the Bayesian optimal design, incorporates a prior distribution for the parameters in the choice model. By accounting for the uncertainties associated with the assumed parameter values, this method provides a better solution compared to a single specification of parameter values. Since its initial proposal by \cite{sandor2001designing}, the Bayesian design approach has found wide application in healthcare, e.g., \cite{Verelst2018}, \cite{deBekkerGrob2019} and \cite{Szinay2021}. 
In this study, we focus on the Bayesian design approach because it demonstrates superior performance compared to other design methods, particularly when there is an abundance of prior information \citep{kessels2006comparison}. 

Thus far, the majority of research on optimal partial profile designs in DCEs has concentrated on main-effects models. Notable contributions to this field include \cite{grasshoff2004optimal}, \cite{kessels2011bayesian, kessels2015improved} and \cite{cuervo2016integrated}. 
%In the main-effects design methods the levels of constant attributes do not matter and are determined randomly. This limits the applicability of the methods to interaction-effects models, for which accurate estimation requires different selections of constant attributes with specified levels. Moreover, in some partial profile designs, the levels of constant attributes are not presented to the respondents, making it impossible to estimate interaction effects. 
However, in the experimental design of DCEs, identifying and accurately estimating interaction effects are widely considered essential \citep{Blomkvist2003}. Although often overlooked by researchers, interaction effects, particularly two-way interactions, frequently occur in practice and are important for understanding subjects’ real-life trade-offs and competing preferences \citep{louviere1990stated,li2013conjoint,jaynes2017minimum}. For example, \cite{johnson2013constructing} highlighted that there is often a significant interaction effect between severity and duration of symptoms. Respondents are unable to evaluate the severity of a migraine without knowing its duration, and similarly, they cannot assess how long the migraine will last without understanding its severity within a specific period. 
Another healthcare intervention DCE, conducted by \cite{Luyten2015} and serving as the motivational example for this paper, revealed that the interaction terms between the type of healthcare intervention and the severity of the disease, as well as between the type of healthcare intervention and the patient's age, significantly influence the public's prioritization of healthcare interventions. Likelihood ratio tests indicated that the relative importance of these interaction terms was even higher than that of some main effects. Neglecting these interaction effects in a choice model may severely diminish its explanatory power \citep{hagerty1986cost}. 

To date, research on partial profile designs based on interaction-effects models remains relatively scarce. Optimal designs are generally model-specific, and designs optimized for main-effects models may not perform well in the context of interaction-effects models \citep{yu2008model}. Some researchers have attempted to modify and adapt partial profile design algorithms developed for main-effects models to make them suitable for interaction-effects models. For instance, in \cite{Luyten2015}, they made slight modifications to the algorithm of \cite{kessels2015improved}, retaining the structure of constant attributes from the main-effects model while computing the optimal levels of these attributes specifically for the full interaction-effects model. 
However, this strategy has three limitations. First, the structure of constant attributes determined by this method has significant room for improvement. One reason for this is that the 
algorithm of \cite{kessels2011bayesian} when identifying constant attributes, focuses primarily on the number of levels for each attribute while overlooking any prior information associated with each attribute. Additionally, this algorithm is sequential in nature: in the first stage, it determines which attributes in each choice set are constant, and in the second stage, it selects the optimal levels for the attributes. This sequential structure often performs worse than algorithms based on simultaneous optimization in complex optimization tasks \citep{vidal2013heuristics, Rashid2017Comparison, Vermetten2020Integrated}. 
Second, in stage two, the algorithm of \cite{kessels2011bayesian} employs a coordinate-exchange (CE) algorithm \citep{meyer1995coordinate,kessels2009efficient} on the master design obtained from stage one to determine the optimal levels of the varying attributes. 
However, the CE algorithm is likely to converge quickly to a local optimum because of its hill-climbing nature, which can affect the quality of the final design obtained \citep{goos2011optimal}.
Third, when determining optimal attribute levels, they base their approach on a full interaction-effects model. In practical applications, however, it is rare for all interactions to be statistically significant, and researchers more commonly retain only significant interactions in their models. As a result, an optimal design derived from a full interaction-effects model may not perform well in the final model, which includes only significant interactions.

To the best of our knowledge, only \cite{grossmann2017partial,grossmann2019practical} and \cite{Manna2024} have discussed optimal partial profile designs for interaction-effects models. However, these works have two limitations. First, their approach is applicable  to specific cases where the choice set comprises only two profiles. Second, their methodology is based on the unrealistic assumption that respondents have equal preferences for all profiles, limiting its applicability to utility-neutral designs, which generally provide less precise parameter estimates than Bayesian optimal designs.

To address these shortcomings, this paper introduces a Simulated Annealing (SA) algorithm based on simultaneous optimization for constructing Bayesian optimal partial-profile interaction-effects designs, without being constrained by the number of profiles or attributes.
This SA algorithm can simultaneously optimize both constant and varying attributes, regardless of whether the model under study is main-effects or interaction-effects. Additionally, this SA algorithm accepts not only better solutions in each iteration, but also accepts worse solutions with a certain probability, allowing it to explore a wider range of possible better partial profile designs. Through extensive simulation experiments, we demonstrate that our SA algorithm can generate better partial profile designs within the same time frame compared to the algorithm of \cite{kessels2015improved}.

%During the experimental design phase, the existence and magnitude of interaction effects are often unknown. In practical data analysis, if interaction effects are found to be insignificant, researchers often resort to employing a main-effects model \citep{yu2008model}. Consequently, this study examines the stability of three different experimental design strategies when faced with inconsistencies between the models used in the experimental design and data analysis phases. Furthermore, a model-robust experimental design strategy is proposed to ensure relatively favorable performances regardless of the model employed during the data analysis phase. Through extensive simulation experiments, we demonstrate that our robust partial profile design consistently performs well against various misspecifications of the underlying model. Additionally, we use a real-life case study to show that our robust partial profile design achieves more precise parameter estimates compared to other experimental design approaches.

In constructing interaction-effects partial profile designs, a practical issue is that the existence and magnitude of interaction effects are often unknown. To mitigate the negative impact of model misspecifications on experimental design, we also applied our SA algorithm to the model-robust designs proposed by \cite{yu2008model}. Through comprehensive simulation experiments, we demonstrate that SA model-robust partial profile designs consistently perform well under various model misspecifications. Additionally, we revisit the healthcare intervention DCE by \cite{Luyten2015} to show that our SA robust partial profile design achieves more precise parameter estimates in practical application compared to other experimental design approaches.

The remainder of this paper is organized as follows. 
In Section \ref{Sec:example}, we use a healthcare intervention DCE as an example to highlight the potential issues present in current partial profile designs based on interaction-effects models.
Section \ref{Sec:Methodology} introduces the multinomial logit (MNL) model for analyzing choice data and discusses how to construct a model-robust Bayesian $\mathcal{D}$-optimal design based on this model. 
In Section \ref{sec:alg}, we propose an SA algorithm for the construction of optimal partial profile designs. 
In Section \ref{sec:CE_SA}, we compare the performance of our integrated SA algorithm with the two-stage CE algorithm by extensive simulation experiments.
In Section \ref{sec:robust design}, we demonstrate how to construct model-robust partial profile designs and compare their estimation performance with that of main-effects and interaction-effects partial profile designs.
In Section \ref{sec:real-life}, we employ our SA algorithm to construct a model-robust design, addressing common challenges encountered in the design of real life DCEs.
Finally, Section \ref{sec:conclusion} summarizes our findings and outlines potential directions for future research.

\section{Motivating Example}\label{Sec:example}
\cite{Luyten2015} conducted a DCE to investigate the public’s distributive preferences regarding healthcare interventions. Their study involved seven attributes, as listed in Table \ref{tab:health_attributes} of Appendix A. Due to the large number of attributes being studied and to reduce participants' cognitive load, they considered a partial profile design where each choice set contains three constant attributes. Besides focusing on the main effects of the attributes, they were also interested in the two-way interactions between $x_1$ and any of the other attributes. However, due to the exclusion of four unrealistic attribute combinations, among which $x_1=$ Curative \& $x_6=$ After 20 years, and $x_1=$ Curative \& $x_6=$ After 5 years, the interaction effects between $x_1$ and $x_6$ could not be estimated. In their work, the authors used effects-type coding for all attributes. Consequently, they needed to estimate 15 main-effects and 12 interaction-effects parameters in total. To provide sufficient information content for the DCE, they constructed a Bayesian $\mathcal{D}$-optimal partial profile design involving 42 choice sets with 2 profiles in each choice set. The 42 choice sets were evenly distributed across three survey groups, each containing 14 choice sets. An example of a choice set appears in Figure \ref{fig:choice_set} of Appendix A. 

The complete details of the partial profile design are displayed in Table \ref{tab:health care org designs} of Appendix A, where the constant attributes in each choice set are highlighted in gray. This design is generated by the two-stage algorithm proposed by \cite{kessels2011bayesian}. In stage one, the algorithm determines which variables should be set as constants in each choice set using an attribute balance method inspired by balanced incomplete block designs (BIBDs) proposed by \cite{green1974}. 
A BIBD arranges $t$ levels of a single qualitative factor, termed treatments, into $S$ blocks. Each block contains $t_v$ treatments, where $t_v < t$. In the context of partial profile design, a choice set can be viewed as a block, and the varying attributes within each choice set can be considered as treatments. BIBDs ensure precise parameter estimates of treatment effects $\alpha_i$ in a two-way ANOVA model. This model can be expressed as
\begin{equation} \label{eq: ANOVA}
    Y_{is} = \alpha_i + \delta_s + \varepsilon_{is},
\end{equation}
where $Y_{is}$ denotes the response of treatment $i$ in block $s$, $\alpha_i$ is the average response of treatment $i$, $\delta_s$ represents the fixed block effects for block $s$, and $\varepsilon_{is}$ is the random error associated with the response of treatment $i$ in block $s$. All random errors are assumed to be independently and normally distributed with a mean of zero and variance $\sigma^2$. In matrix notation, the two-way ANOVA model can be expressed as
\begin{equation}\label{eq: ANOVA_matrix}
    \boldsymbol{Y} = \boldsymbol{Q} \boldsymbol{\alpha} + \boldsymbol{Z} \boldsymbol{\delta} + \boldsymbol{\varepsilon},
\end{equation}
where $\boldsymbol{Y}$ is a vector of dimension $r = S \times t_v$ representing the responses, $\boldsymbol{Q}$ is an $r \times t$ design matrix for the treatments, $\boldsymbol{\alpha} = (\alpha_1,\dots,\alpha_t)^T$ represents the mean responses for each treatment, $\boldsymbol{Z}$ is an $r \times (S-1)$ design matrix hat maps treatments to blocks, $\boldsymbol{\delta} = (\delta_1,\dots,\delta_{S-1})^T$ is the vector of block effects, and $\boldsymbol{\varepsilon}$ represents the random error vector. 

BIBDs only exist for certain specified combinations of $S$, $t$, and $t_v$, which limits their applicability in practical settings. To address this, \cite{kessels2011bayesian} proposed constructing $\mathcal{D}$-optimal designs for the two-way ANOVA model defined in Eq. (\ref{eq: ANOVA_matrix}) by maximizing the determinant of the information matrix $\boldsymbol{M}$ for the parameter $\boldsymbol{\alpha}$ and $\boldsymbol{\delta}$, given by:
\begin{equation} 
    \boldsymbol{M} = \begin{bmatrix}
    \boldsymbol{Q}^T\boldsymbol{Q}&\boldsymbol{Q}^T\boldsymbol{Z}\\
\boldsymbol{Z}^T\boldsymbol{Q}&\boldsymbol{Z}^T\boldsymbol{Z}\end{bmatrix}.
\end{equation}

In the master design obtained using the $\mathcal{D}$-optimality criterion, equal attention is paid to each attribute, which is why this method is called the attribute balance approach. However, in practice, attributes often have different numbers of levels. Attributes with more levels typically need to be assigned as varying attributes more frequently to achieve equally precise part-worth estimates as those with fewer levels. 

To address this, \cite{kessels2015improved} improved the algorithm by introducing two variance balance approaches that construct weighted $\mathcal{A}$-optimal designs for the ANOVA model defined in Eq.~(\ref{eq: ANOVA}). 
The weighted $\mathcal{A}$-optimal criterion is adopted instead of the $\mathcal{D}$-optimal criterion because it minimizes the weighted sum of the parameter variances, thereby directly targeting the balance of estimation precision across attributes with heterogeneous numbers of levels. 
In contrast, $\mathcal{D}$-optimal (attribute balance) designs treat all attributes equally and are only appropriate when all attributes have the same number of levels. 

The weighted $\mathcal{A}$-optimal design seeks to minimize
\begin{equation}
    \mathcal{A}_w = \sum_{i=1}^{t} w_i \text{Var}(\hat{\alpha}_i),
\end{equation}
where $w_i$ is the weight assigned to the variance of the $i$-th individual parameter estimator $\hat{\alpha}_i$ in the two-way ANOVA model. The variances are given by the diagonal elements of the upper left-hand submatrix of the inverse of the information matrix $\boldsymbol{M}^{-1}$, which can be calculated as
\begin{equation}
    \left\{ \boldsymbol{Q}^T \left( \boldsymbol{I}_r - \boldsymbol{Z} \left( \boldsymbol{Z}^T \boldsymbol{Z} \right)^{-1} \boldsymbol{Z}^T \right) \boldsymbol{Q} \right\}^{-1},
\end{equation}
where $\boldsymbol{I}_r$ is an $r \times r$ identity matrix.

In an ANOVA model, a larger $w_i$ indicates that the treatment effect $\alpha_i$ is expected to be estimated more precisely than other treatment effects. In a partial profile design, let $d_i$ represent the number of levels for the $i$-th attribute. A larger $d_i$ suggests that this attribute requires more information for accurate estimation, and thus its corresponding weight should also be larger. 
\cite{kessels2015improved} introduces two methods to calculate these weights to construct an $\mathcal{A}_w$-optimal master design, referring to these methods as variance balance I and variance balance II. 

In the variance balance I approach, the weight $w_i$ is defined as the ratio of the number of part-worth values for attribute $i$ to the total number of part-worth values:
\begin{equation}
  w_i = \frac{d_i - 1}{\sum_{i=1}^t d_i - t}.
\end{equation}
Variance balance I therefore increases the weight linearly with the number of levels, ensuring that the loss of information due to constant attributes is distributed evenly across all part-worths.

In the variance balance II approach, the weight $w_i$ can be calculated as
\begin{equation}\label{eq:var_balance_2}
    w_i = \frac{(d_i - 1)^2}{2d_i}.
\end{equation}
This formula is derived from the information matrix of utility-neutral optimal full-profile designs \citep{GRAHOFF2004361}. 
Compared with variance balance I, variance balance II enlarges the weight differences across attributes, giving proportionally greater weight to attributes with more levels or higher complexity. 
Hence, variance balance II is preferred when the goal is to achieve more precise estimation for attributes with many levels.

After obtaining the master design generated in stage one, \cite{kessels2011bayesian,kessels2015improved} use a restricted CE algorithm to determine the optimal levels of the attributes.
The CE algorithm initiates by creating a random initial design as a starting point.
The algorithm iteratively optimizes this initial design by evaluating each attribute level across profiles. 
For each profile, the algorithm examines every attribute individually, calculating the optimality criterion for each level of that attribute. 
A level substitution is applied only if it results in an improved criterion value. 
This attribute-by-attribute examination continues across all profiles, completing one full cycle when the algorithm has identified the best possible exchange for each attribute in every profile. If any updates are made during this cycle, another complete iteration begins from the first attribute of the first profile. The process repeats until a full cycle is completed without any changes or until a specified maximum number of iterations is reached.

In partial profile design, \cite{kessels2011bayesian,kessels2015improved} introduces an additional constraint to the CE algorithm's iterative process, specifically restricting the levels of the non-constant attributes to vary within each choice set. This constraint prevents the occurrence of choice sets in the resulting optimal design where the number of varying attributes falls below $t_v$, characterizing it as the restricted CE algorithm.

%is an incomplete block design where all pairs of treatments occur together within a block an equal number of times. Based on the master design obtained in stage one, they use the CE algorithm in stage two to determine the varying attribute levels. They later improved this algorithm in \cite{kessels2015improved} by applying variance balance methods in stage one to determine the varying attributes. In contrast to the earlier attribute balance method, variance balance approaches take into account the existence of different numbers of levels for each attribute. To ensure similar statistical information for each part-worth estimate, attributes with more levels need to be fixed less frequently.

Although the improved two-stage algorithm accounts for heterogeneity in the number of levels of each attribute, it is not well developed for interaction-effects models. 
This limitation arises because, in interaction-effects models, even attributes with the same number of levels can be involved in different interactions. \cite{Zwerina1996AGM} emphasized that the presence of interaction effects requires overlap in attribute levels within choice sets to generate the necessary contrasts for estimation. Therefore, if an attribute appears in multiple interaction terms within the model, it should remain constant more frequently to facilitate the estimation of these interaction effects.
Consequently, considering only the differences in the number of levels among attributes is insufficient for accurately determining how fixed attributes should be distributed within choice sets.
Moreover, the issue is further exacerbated by the fact that, in the two-stage CE algorithm, the master design determined in stage one remains fixed in stage two. This constraint prevents any adjustments, even if the initially generated master design proves to be inefficient for estimating interaction effects within the model.

To address this, we propose a simultaneous-optimization SA algorithm to handle this complex problem. 
Unlike the two-stage algorithm, our SA algorithm use the Bayesian $\mathcal{D}_B$-optimality criterion to simultaneously optimize the allocation of constant attributes and the levels of all attributes, ensuring greater flexibility across different model under study.
Moreover, as pointed out by \cite{MAO2025105305}, compared to the hill-climbing CE algorithm, the SA algorithm is more likely to escape local optima and explore a broader experimental region, thus generating superior choice designs within a given time frame.

Beyond the algorithm, this healthcare DCE also encountered a common issue in the optimal design of interaction-effects models: the misspecification of the underlying model. In the experimental design phase, they consider a MNL model with 12 interaction effects. Subsequent data analysis, however, indicated that only the interaction terms between $x_1$ and $x_4$ as well as $x_1$ and $x_7$ were significant. Consequently, they dropped the non-significant interaction terms in the final model. It is important to note that optimal designs generated for a specific model do not necessarily perform well when applied to other models \citep{Atkinson1996}. During the experimental design phase, the limited information about interactions complicates the accurate identification of the true underlying model. To address this, \cite{yu2008model} proposed a robust strategy for full profile designs that performs well even in the face of model misspecification. In this paper, we extend this strategy to partial profile designs and employ our SA algorithm to construct such model-robust designs.
 
\section{Methodology}\label{Sec:Methodology}
\subsection{Multinomial Logit Model Bayesian $\mathcal{D}$-Optimality}
The MNL model is based on the assumption of utility maximization, which means that participants in the experiment will choose the profile with the highest utility in each choice set. Suppose each participant faces $S$ different choice sets, with each choice set containing $J$ distinct profiles. The perceived utility $U_{sj}$ of a respondent choosing profile $j$ in choice set $s$ can be expressed as
\begin{equation}
    U_{sj}= \boldsymbol{x}_{sj}^{T}\boldsymbol{\beta}+ \varepsilon_{sj},\label{eq：utility}
\end{equation}
where $\boldsymbol{x}_{sj}$ represents the attribute vector, $\boldsymbol{\beta}$ is the vector of corresponding coefficients. The MNL model assumes that the random errors $\varepsilon_{sj}$ are independently and identically distributed according to a Gumbel (or Type I extreme value) distribution. And therefore, the MNL probability that $p_{js}$ profile $j$ is selected in choice set $s$ can be expressed as
\begin{equation}
p_{js} = \frac{\mbox{exp}\left({\boldsymbol x}^{T}_{js}\boldsymbol{\beta}\right)}{\sum_{j=1}^{J}\mbox{exp}\left({\boldsymbol x}^{T}_{js}\boldsymbol{\beta}\right)}. \label{eq: secondmnl}
\end{equation}
The coefficient vector $\boldsymbol{\beta}$ can be estimated by maximizing the log-likelihood function:
\begin{equation}\label{eq:log-likelihood}
LL(\boldsymbol{\beta})=\sum_{n=1}^{M}\sum_{s=1}^{S}\sum_{j=1}^{J} y_{nsj}\ln(p_{nsj}),
\end{equation}
where $y_{nsj}$ is a binary variable which equals one if respondent $n$ chooses profile $j$ from choice set $s$ and zero otherwise.
To construct optimal choice designs, many researchers employ the $\mathcal{D}$-optimality criterion \citep{sandor2001designing}, which leads to a $\mathcal{D}$-optimal design. The $\mathcal{D}$-optimal design obtains more precise parameter estimates by maximizing the determinant of the information matrix. The $\mathcal{D}$-optimal optimality criterion is defined as 
\begin{equation}
\mathcal{D} = \mbox{log} \left|{\bf M}\left({\bf X}, \boldsymbol{\beta}\right)\right|, \label{eq:Dcriterion}
\end{equation}
where $\bf{X}$ is the design matrix and ${\bf M}\left({\bf X}, \boldsymbol{\beta}\right)$ is the information matrix of the model under study. In MNL models, the information matrix ${\bf M}\left({\bf X}, \boldsymbol{\beta}\right)$ can be calculated as the sum of the information matrix of each chocie set $s$, that is, 
\begin{equation}
{\bf M}\left({\bf X}, \boldsymbol{\beta}\right) = \sum_{s=1}^S {\bf X}^{T}_{s}\left({\bf P}_{s} - {\bf p}_{s}{\bf p}^{T}_{s}\right){\bf X}_{s}, \label{info}
\end{equation}
where  $\bf{X}=({\mathbf X}_{1},\dots,{\mathbf X}_{s})$ is the total model matrix for all $S$ choice sets, ${\bf P}_{s} = \mbox{diag}\left({\bf p}_{s}\right)$ with ${\bf p}_{s} = \left(p_{1s},\dots, p_{Js}\right)^{T}$ records the MNL probabilities of all $J$ profiles in choice set $s$.

According to Eq. (\ref{eq: secondmnl}), the MNL probabilities are dependent on the parameter vector $\boldsymbol{\beta}$, which is often unknown during the experimental design phase. To address this issue, researchers often assume that the parameter vector $\boldsymbol{\beta}$ follows a certain prior distribution, leading to a Bayesian optimal design approach \citep{kessels2011bayesian}. Denote the prior distribution for $\boldsymbol{\beta}$ as $\pi(\boldsymbol{\beta})$, the Bayesian $\mathcal{D}$-optimality criterion seeks to maximize the determinant of the information matrix averaged over $\pi(\boldsymbol{\beta})$, that is, 
\begin{equation}
\mathcal{D}_B = \int_{\mathcal{R}^m} \mbox{log} \left|{\bf M}\left({\bf X}, \boldsymbol{\beta}\right)\right| \pi(\boldsymbol{\beta})\mbox{d}\boldsymbol{\beta}.\label{eq:Dbcriterion}
\end{equation}
Given that Eq. (\ref{eq:Dbcriterion}) does not have a closed-form solution, the $\mathcal{D}_B$-optimality criterion is often numerically approximated through sampling from the prior distribution $\pi(\boldsymbol{\beta})$. In our work, we utilize the spherical-radial transformation sampling method proposed by \cite{gotwalt2009fast}, which shows superior efficacy of in evaluating the Bayesian optimality criterion over other approaches \citep{yu2010comparing}.

To compare the statistical efficiency of different partial profile designs, we use the relative $\mathcal{D}_B$-efficiency as a measure \citep{holling2011usefulness}. The Bayesian $\mathcal{D}_B$-efficiency of a design $\mathbf{X}$ relative to another design $\mathbf X^*$ can be calculated as
\begin{equation}\label{eq:D-efficiency}
    \mathcal{D}_{B}\text{-eff}(\mathbf{X},\mathbf{X^*}) = \exp\left(\frac{\mathcal{D}_B(\mathbf{X})-\mathcal{D}_B(\mathbf{X^*})}{\textit{m}}\right),
\end{equation}
where $m$ is the dimension of the parameter vector $\boldsymbol{\beta}$.
A value of 1 indicates equal statistical efficiency for the two designs. A value greater than 1 indicates that the design $\mathbf{X}$ performs better and vice versa.

\subsection{Model Robust Bayesian $\mathcal{D}$-Optimality}
Due to the limited information about interaction effects during the experimental design phase, it is challenging to determine the underlying model. An optimal experimental design for a specific model may be inefficient in alternative models \citep{Atkinson1996}. Consequently, we introduce a model-robust experimental design strategy that ensures our design performs relatively well even faced with model misspecification.

Let $\boldsymbol{\beta}_{main}$ and $\boldsymbol{\beta}_{int}$ be the parameter vectors associated with these models, with dimensions $m_{main}$ and $m_{int}$, respectively. According to the definition of the $\mathcal{D}$-optimality criterion in Eq. (\ref{eq:D main}), we have:
\begin{equation}
\mathcal{D}_{main} = \mbox{log} \left|{\bf M}\left({\bf X}, \boldsymbol{\beta}_{main}\right)\right|,\label{eq:D main}
\end{equation}
and
\begin{equation}
\mathcal{D}_{int} = \mbox{log} \left|{\bf M}\left({\bf X}, \boldsymbol{\beta}_{int}\right)\right|.\label{eq:D int}
\end{equation}
As in \cite{yu2008model}, our model-robust design use a composite design criterion based on both the main-effects and interaction-effects model, that is, 
\begin{equation}
\mathcal{D}_{robust} =\frac{\mathcal{D}_{main}}{m_{main}}+ \frac{\mathcal{D}_{int}}{m_{int}}.\label{eq:D robust}
\end{equation}
Similarly, the Bayesian composite design criterion can also be calculated as the average over the parameter distribution.
\section{Simulated Annealing Algorithm}\label{sec:alg}

Due to the nonlinear nature of the MNL model, constructing optimal choice designs often necessitates the use of a search algorithm. Currently, most algorithms used to construct Bayesian optimal experimental designs are based on hill-climbing approaches, such as the CE algorithm \citep{meyer1995coordinate,kessels2006comparison}. These algorithms, during iterations, accept only improved choice designs, thus leading to rapid convergence. To address this issue, \cite{MAO2025100551} has developed an SA algorithm for constructing Bayesian optimal designs for DCEs. Since the work of \cite{MAO2025100551} focuses on full profile designs and could not be applied to partial profile designs, we have made several modifications based on their approach, resulting in the algorithm defined in Algorithm \ref{alg:the_alg}.
\begin{algorithm}[h]
    \caption{Pseudo code for Simulated Annealing}
    \label{alg:the_alg}
    \SetAlgoLined
    \SetKwInOut{Input}{Input}
    \SetKwInOut{Output}{Output}
    \Input{Initial random partial profile design $\mathbf{X}$}
    \Output{The best partial profile design $\mathbf{X}_{Best}$ found by the algorithm}
    Set a value for \textit{Initial Temperature} $T_0$ by a random walk approach\;
    Set iteration counter $k = 0$\;
    Set $\mathbf{X}_{Best} = \mathbf{X}$\;
    \While{Stopping Criterion not met}{
        Generate the new design $\mathbf{X'}$ following the \textit{Exploration Rule} defined in Algorithm \ref{alg:exploration_rule}\;
        Let $\mathcal{D}_B(\mathbf{X})$ be the Bayesian $\mathcal{D}$-optimality criterion for the partial profile $\mathbf{X}$. Accept $\mathbf{X'}$ as the new solution with probability $p$ such that:
        \begin{equation}
            p(\mathbf{X}, \mathbf{X'}) = \min\left\{1, \exp\left(\frac{\mathcal{D}_B(\mathbf{X'}) - \mathcal{D}_B(\mathbf{X})}{T_{k}}\right)\right\}\label{eq:prob}
        \end{equation}\\
        \If{$\mathbf{X'}$ is accepted}{
            Set $\mathbf{X} = \mathbf{X'}$\;
            \If{$\mathcal{D}_B(\mathbf{X}) > \mathcal{D}_B(\mathbf{X}_{Best})$}{
                $\mathbf{X}_{Best} = \mathbf{X}$\;
            }
        }
        $k = k + 1$\;
        \If{no new solutions are accepted in the last 1000 iterations}{
        Reheat the temperature: $T_{k} = T_0$\;
        Reset iteration counter: $k = 0$\;
        }

    }
    \textbf{return} $\mathbf{X}_{Best}$
\end{algorithm}

The SA algorithm is a probabilistic technique used to find or approximate the global optimum of a given objective function by simulating the process of slowly cooling a material to increase the stability of the system.
This algorithm requires a random initial partial profile design $\mathbf{X}$ as input. It begins by generating a problem-specific initial temperature $T_0$ through a random walk method as in \cite{MAO2025100551}. Together with $T_0$, the iteration counter $k$ is set to zero, and the initial design $\mathbf{X}$ is considered as the current best design $\mathbf{X}_{Best}$. In each iteration, a specific cooling function $f(k,T_0) $ is used to gradually reduce the system temperature. In our work, we use a Hyperbolic cooling function $ f(k,T_0) = \frac{T_0}{k+1}$. 
Through extensive experiments, \cite{MAO2025100551} has demonstrated that this cooling function effectively balances design quality and computational efficiency in the context of Bayesian optimal experimental design for DCEs.
Simultaneously, a new random partial profile design $\mathbf{X'}$ is generated in the neighborhood of $\mathbf{X}$ based on the exploration rule as defined in Algorithm \ref{alg:exploration_rule}. Once $\mathbf{X'}$ is generated, its Bayesian $\mathcal{D}$-optimality criterion is calculated, and the Metropolis acceptance criterion \citep{metropolis1953equation}, as defined in Eq. (\ref{eq:prob}), is used to decide whether to accept this new design. Unlike hill-climbing algorithms, this algorithm accepts not only all superior solutions but also, with a certain probability, inferior ones. If the newly introduced solution is better than the current optimum, $\mathbf{X}_{Best}$ is updated accordingly. As the algorithm progresses, the decrease in the temperature of the system gradually reduces the likelihood of accepting poorer solutions. At very low temperatures, the algorithm behaves similarly to a hill-climbing approach, primarily accepting only superior solutions. To prevent premature convergence, if no new solutions are accepted within 1000 iterations, the system temperature is reheated to $T_0$, enabling exploration of more potential solutions.  
This process iterates continuously until the predetermined stopping criterion is met. Once the stopping criterion is satisfied, the algorithm terminates and outputs $\mathbf{X}_{Best}$ as the optimal partial profile design. 
Typically, the stopping criterion can be defined as the algorithm running until either a predetermined maximum runtime is reached or the number of reheating cycles reaches a specified limit. Alternatively, a more adaptive criterion can be used, such as terminating the algorithm when no new best solution is found within an entire reheating cycle.

Compared to the work of \cite{MAO2025100551}, the primary distinction in our approach lies in the modification of the exploration rule. In their work, modifications to $\mathbf{X}$ were made by randomly altering the level of a randomly selected attribute within a randomly chosen choice set. However, this method does not address the constraint in constructing partial profile designs where each choice set must maintain a fixed number of attributes as constant. 

\begin{algorithm}[h]
    \caption{Pseudo code for Exploration Rule}
    \label{alg:exploration_rule}
    \SetAlgoLined
    \SetKwInOut{Input}{Input}
    \SetKwInOut{Output}{Output}
    \Input{Current partial profile design $\mathbf{X}$ and a threshold $\gamma$}
    \Output{Updated partial profile design $\mathbf{X'}$}
    Randomly select an attribute $i$ from profile $j$ in choice set $s$ of $\mathbf{X}$\;

    \If{attribute $i$ is constant}{
        \If{$\mathcal{D}_B(\mathbf{X})$ is independent with the level of constant attribute $i$ }{
            Randomly select a varying attribute in $s$, convert it to a constant attribute with a random level\;
            Randomly change the level of attribute $i$ for profile $j$ in $s$\;
        }
        \Else{
            Generate a random probability $p$\;
            \If{$p \leq \gamma$}{
                Randomly modify the level of the constant attribute $i$\;
            }
            \Else{
               Randomly select a varying attribute in $s$, convert it to a constant attribute with a random level\;
               Randomly change the level of attribute $i$ for profile $j$ in $s$;
            }
        }
    }
    \Else{
        Modify the level of attribute $i$ for profile $j$ in $s$\;
        \If{attribute $i$ becomes a constant attribute}{
            Randomly select a varying attribute in $s$, convert it to a constant attribute with a random level\;
        }
    }
\end{algorithm}

To address this limitation, we utilized the exploration rule defined in Algorithm \ref{alg:exploration_rule}.  Our exploration rule begins by selecting a random attribute $i$ from a randomly chosen profile $j$ within a random choice set $s$ of the current design $\mathbf{X}$. 
We then check whether the attribute $i$ is constant. 
If so, we assess whether changing the level of this constant attribute affects the Bayesian $\mathcal{D}$-optimality criterion, specifically to rule out:
\begin{enumerate}
\item Attribute $i$ is not involved in any interaction effects that we aim to study.
\item All interaction effects involving this attribute in our model are comprised exclusively of other constant attributes from current choice set $s$.
\end{enumerate}
If either of these conditions holds, changing the level of this constant attribute does not impact the information matrix.
In such cases, we proceed by converting this constant attribute into a varying one and randomly selecting another varying attribute to be fixed at a random level, thereby maintaining the required structure of the partial profile design.
Conversely, if modifying the attribute $i$ affects the information matrix, we determine whether to alter its level based on a predetermined threshold $\gamma$.
This threshold, ranging between 0 and 1, correlates directly with the probability of changing the level of a constant attribute. The higher the threshold, the greater the likelihood.In this study, we define $\gamma$ as the ratio of constant attributes to the total number of attributes. 
If the selected attribute is not constant, we randomly modify its level. It is important to note that this operation may result in the attribute transforming from varying to constant. To maintain constant profile strength, under such circumstances, we convert a randomly selected constant attribute to varying. Our approach carefully accounts for the inherent constraints of partial profile designs while ensuring minimal perturbation to the existing design. This strategy balances structural constraints with sufficient exploration, allowing the algorithm to efficiently explore the design space during optimization.

\section{Algorithm Comparison}\label{sec:CE_SA}
In this section, we evaluate the performance of our SA algorithm and the algorithm of \cite{kessels2015improved} through computational experiments.
All code used in this study was developed in in MATLAB R2024b and executed on a system powered by an 11th Gen Intel(R) Core(TM) i5-1135G7 processor.

\subsection{Simulation setup}\label{subsec:setup}
To control the size of the experiment, we consider DCEs with 24 choice sets and 6 attributes, with attribute levels of 2, 2, 2, 3, 3, and 3, respectively. To obtain generalized conclusions, we consider various experimental settings defined by the number of profiles ($J$), the number of constant attributes ($F$), the number of two-way interaction terms ($Int$), and the prior distributions.

 For the number of profiles, we consider scenarios with 2 or 3 profiles. For the number of constant attributes, we consider scenarios with 1 or 2 constant attributes. Regarding the number of interaction terms, we consider four scenarios: no interaction terms, interactions between the first attribute and the other 2-level attributes, interactions between the first attribute and all 3-level attributes, and interactions between the first attribute and all other attributes. In our work, we employ effects-type coding for all attributes, ensuring that the sum of the levels for each attribute equals zero, which necessitates the estimation of coefficients for all levels of each attribute excluding the last level. Correspondingly, the number of interaction terms is 0, 2, 6, and 8, respectively.

 The prior distribution includes the prior mean $\boldsymbol{\beta}_0$ and prior variance $\boldsymbol{\Sigma}_0$. Regarding the prior mean $\boldsymbol{\beta}_0^{main}$ for main effects, we assume that preferences for each attribute increase with the level, meaning that the first level is least favored and the last level is most favored. 
 For generality, we define $\boldsymbol{\beta}_0^{main}$ as a function of a scale parameter $\lambda$:

 \begin{equation}
    \boldsymbol{\beta}_{0}^{main}(\lambda)=(-\lambda, -\lambda, -\lambda, -\lambda, 0, -\lambda, 0,-\lambda, 0)^{T},
\end{equation} 
where $\lambda \in \{1,\frac{1}{2},\frac{1}{3}\}$. Similarly, we set the prior covariance matrix $\boldsymbol{\Sigma}_0^{main}$ of the main effects as a function of a scale parameter $\kappa$:
\begin{equation}
    \boldsymbol{\Sigma}_0^{main}(\kappa) = \begin{pmatrix}
\kappa^2 & 0 & 0 & 0 & 0 & 0 & 0 & 0 & 0 \\
0 & \kappa^2 & 0 & 0 & 0 & 0 & 0 & 0 & 0 \\
0 & 0 & \kappa^2 & 0 & 0 & 0 & 0 & 0 & 0 \\
0 & 0 & 0 & \kappa^2 & -0.5\kappa^2 & 0 & 0 & 0 & 0 \\
0 & 0 & 0 & -0.5\kappa^2 & \kappa^2 & 0 & 0 & 0 & 0 \\
0 & 0 & 0 & 0 & 0 & \kappa^2 & -0.5\kappa^2 & 0 & 0 \\
0 & 0 & 0 & 0 & 0 & -0.5\kappa^2 & \kappa^2 & 0 & 0 \\
0 & 0 & 0 & 0 & 0 & 0 & 0 & \kappa^2 & -0.5\kappa^2 \\
0 & 0 & 0 & 0 & 0 & 0 & 0 & -0.5\kappa^2 & \kappa^2 \\
\end{pmatrix},
\end{equation}
where $\kappa \in \{1,\frac{1}{2},\frac{1}{3}\}$.
For attributes with three levels, we ensure a negative correlation between the parameters of the first two levels, thus maintaining equal variances across all parameters, particularly for the third level \citep{kessels2008recommendations}.
By combining different values of $\lambda$ and $\kappa$, we obtain a total of nine distinct prior distributions for the main effect parameters, broadly covering most common scenarios encountered in practice.

Regarding the prior distribution for interaction effects, we apply a naive prior, setting the prior mean to zero and the variance to one. This choice corresponds with the typically limited information available on interaction effects in most practical scenarios. For further discussion on the strategies of setting prior information for interaction effects, please refer to Section \ref{sec:robust design}. 
 
Utilizing distinct levels of these five key parameters that define the DCE settings, we implemented a full factorial experiment encompassing 2×2×4×3×3 = 144 different DCE scenarios. In each scenario, we constructed Bayesian optimal experimental designs using both the two-stage CE algorithm of \cite{kessels2015improved} and our SA algorithm.

To ensure a fair comparison, we determined the runtime of both algorithms following the approach used in \cite{MAO2025105305}.
Specifically, for each scenario, we first run the CE algorithm, record its total runtime, and use this as the stopping criterion for the SA algorithm, setting its maximum allowable runtime accordingly.

Regarding the CE algorithm's parameter settings, in stage one, we apply the variance balance II approach, as introduced in Eq. (\ref{eq:var_balance_2}), to generate the master design. In stage two, we initialize the algorithm with 30 random starting points.
To guarantee full convergence of the CE algorithm, we terminate it at each starting point when its $\mathcal{D}_{B}$-value shows no improvement over an entire cycle. Finally, among the 30 starting points, the best partial profile design obtained is selected as the final output of the CE algorithm.

\subsection{Results}\label{subsec: result}
In each experimental setting, we documented the final partial profile designs ${\bf {X}}_{CE}$ and ${\bf {X}}_{SA}$ generated by these two algorithms and calculated the relative $\mathcal{D}_{B}$-efficiency of the CE designs compared to the SA designs labeled as $\mathcal{D}_{B}\text{-eff}$, which are detailed in Table \ref{tab:CE_SA}. 
The results in the table indicate that the $\mathcal{D}_{B}\text{-eff}$ is less than 1 in all 144 scenarios, demonstrating the superior performance of the SA designs. 
On average, the relative $\mathcal{D}_{B}$-efficiency across all 144 scenarios is about 92.60\%.
We further conducted a Wilcoxon signed-rank test and found that the median of $\mathcal{D}_{B}\text{-eff}$ was significantly less than 1, with a p-value less than 0.0001.

Among the five factors that define the DCE scenarios, two have a particularly strong impact on the relative performance of the algorithms: the size of the prior variance $\kappa$ and the number of two-way interaction terms $Int$.
Regarding $\kappa$, our findings align with those of \citep{MAO2025100551}, showing that the advantage of the SA algorithm tends to be more pronounced as $\kappa$ increases. This is because larger prior variances indicate greater uncertainty in the prior information, making the optimization objective function more complex and increasing the likelihood of encountering multiple local optima. In such cases, the SA algorithm’s ability to accept inferior solutions during the search process allows it to escape local optima more effectively, explore a broader design space, and ultimately identify superior designs.

As for $Int$, the relative performance of the SA algorithm generally improves as the number of interaction terms in the model increases. This is primarily because the first stage of the two-stage CE algorithm does not account for interaction effects. As the influence of interaction effects in the model grows, this omission becomes more consequential, leading to a greater discrepancy between the initially generated master design and the structure of the optimal design. Consequently, the integrated SA algorithm, which does not rely on a predetermined master design, is better suited to adapt to the complexities introduced by interaction effects, thereby achieving superior performance.

\begin{longtable}{ccccccc|ccccccc}
\caption{Relative $\mathcal{D}_B$-efficiency of the CE designs compared to the SA designs based on different experimental settings.}
\label{tab:CE_SA}\\
\hline
$J$ & $F$ & $Int$ & $\lambda$ & $\kappa$ & $\mathcal{D}_{B}\text{-eff}$ & Runtime (s) & $J$ & $F$ & $Int$ & $\lambda$ & $\kappa$ & $\mathcal{D}_{B}\text{-eff}$ & Runtime (s) \\
\hline
2 & 1 & 0 & 1 & 1 & 94.26\% & 534.96   & 3 & 1 & 0 & 1 & 1 & 92.97\% & 599.64   \\
2 & 1 & 0 & 1 & 0.5 & 95.51\% & 431.69   & 3 & 1 & 0 & 1 & 0.5 & 96.28\% & 580.90   \\
2 & 1 & 0 & 1 & 0.33 & 95.57\% & 419.73   & 3 & 1 & 0 & 1 & 0.33 & 95.98\% & 598.21   \\
2 & 1 & 0 & 0.5 & 1 & 94.77\% & 489.28   & 3 & 1 & 0 & 0.5 & 1 & 94.57\% & 588.42   \\
2 & 1 & 0 & 0.5 & 0.5 & 97.76\% & 430.61   & 3 & 1 & 0 & 0.5 & 0.5 & 95.38\% & 585.25   \\
2 & 1 & 0 & 0.5 & 0.33 & 96.13\% & 409.03   & 3 & 1 & 0 & 0.5 & 0.33 & 93.97\% & 561.61   \\
2 & 1 & 0 & 0.33 & 1 & 94.71\% & 515.55   & 3 & 1 & 0 & 0.33 & 1 & 95.69\% & 625.68   \\
2 & 1 & 0 & 0.33 & 0.5 & 96.14\% & 443.94   & 3 & 1 & 0 & 0.33 & 0.5 & 95.92\% & 547.13   \\
2 & 1 & 0 & 0.33 & 0.33 & 96.97\% & 420.64   & 3 & 1 & 0 & 0.33 & 0.33 & 95.28\% & 535.33   \\
2 & 1 & 2 & 1 & 1 & 91.35\% & 774.73   & 3 & 1 & 2 & 1 & 1 & 89.02\% & 998.45   \\
2 & 1 & 2 & 1 & 0.5 & 96.35\% & 745.11   & 3 & 1 & 2 & 1 & 0.5 & 94.61\% & 813.16   \\
2 & 1 & 2 & 1 & 0.33 & 97.77\% & 636.07   & 3 & 1 & 2 & 1 & 0.33 & 95.17\% & 868.89   \\
2 & 1 & 2 & 0.5 & 1 & 89.23\% & 813.96   & 3 & 1 & 2 & 0.5 & 1 & 91.44\% & 990.33   \\
2 & 1 & 2 & 0.5 & 0.5 & 97.37\% & 773.28   & 3 & 1 & 2 & 0.5 & 0.5 & 94.80\% & 957.95   \\
2 & 1 & 2 & 0.5 & 0.33 & 98.22\% & 689.98   & 3 & 1 & 2 & 0.5 & 0.33 & 95.46\% & 936.71   \\
2 & 1 & 2 & 0.33 & 1 & 90.22\% & 923.66   & 3 & 1 & 2 & 0.33 & 1 & 89.38\% & 961.40   \\
2 & 1 & 2 & 0.33 & 0.5 & 96.83\% & 776.80   & 3 & 1 & 2 & 0.33 & 0.5 & 95.30\% & 855.32   \\
2 & 1 & 2 & 0.33 & 0.33 & 99.21\% & 731.94   & 3 & 1 & 2 & 0.33 & 0.33 & 95.08\% & 951.75   \\
2 & 1 & 6 & 1 & 1 & 86.14\% & 2578.77  & 3 & 1 & 6 & 1 & 1 & 86.54\% & 2085.88  \\
2 & 1 & 6 & 1 & 0.5 & 94.64\% & 2011.19  & 3 & 1 & 6 & 1 & 0.5 & 90.77\% & 1681.61  \\
2 & 1 & 6 & 1 & 0.33 & 95.20\% & 2294.73  & 3 & 1 & 6 & 1 & 0.33 & 91.78\% & 1573.64  \\
2 & 1 & 6 & 0.5 & 1 & 89.93\% & 1712.03  & 3 & 1 & 6 & 0.5 & 1 & 85.52\% & 2134.28  \\
2 & 1 & 6 & 0.5 & 0.5 & 94.05\% & 1364.89  & 3 & 1 & 6 & 0.5 & 0.5 & 93.26\% & 1859.44  \\
2 & 1 & 6 & 0.5 & 0.33 & 96.45\% & 1458.95  & 3 & 1 & 6 & 0.5 & 0.33 & 94.84\% & 2085.58  \\
2 & 1 & 6 & 0.33 & 1 & 86.52\% & 1822.05  & 3 & 1 & 6 & 0.33 & 1 & 87.43\% & 2128.47  \\
2 & 1 & 6 & 0.33 & 0.5 & 93.97\% & 1365.48  & 3 & 1 & 6 & 0.33 & 0.5 & 93.65\% & 1714.35  \\
2 & 1 & 6 & 0.33 & 0.33 & 96.37\% & 1432.95  & 3 & 1 & 6 & 0.33 & 0.33 & 94.84\% & 1490.76  \\
2 & 1 & 8 & 1 & 1 & 87.37\% & 3700.49  & 3 & 1 & 8 & 1 & 1 & 83.62\% & 3301.11  \\
2 & 1 & 8 & 1 & 0.5 & 94.95\% & 2604.04  & 3 & 1 & 8 & 1 & 0.5 & 91.21\% & 2896.04  \\
2 & 1 & 8 & 1 & 0.33 & 93.18\% & 2812.72  & 3 & 1 & 8 & 1 & 0.33 & 91.41\% & 2698.85  \\
2 & 1 & 8 & 0.5 & 1 & 86.91\% & 3666.94  & 3 & 1 & 8 & 0.5 & 1 & 83.00\% & 3237.76  \\
2 & 1 & 8 & 0.5 & 0.5 & 97.60\% & 2760.64  & 3 & 1 & 8 & 0.5 & 0.5 & 91.36\% & 2759.62  \\
2 & 1 & 8 & 0.5 & 0.33 & 98.77\% & 2621.92  & 3 & 1 & 8 & 0.5 & 0.33 & 93.66\% & 2830.41  \\
2 & 1 & 8 & 0.33 & 1 & 86.45\% & 3704.42  & 3 & 1 & 8 & 0.33 & 1 & 82.80\% & 3783.41  \\
2 & 1 & 8 & 0.33 & 0.5 & 97.01\% & 2338.69  & 3 & 1 & 8 & 0.33 & 0.5 & 92.68\% & 2958.64  \\
2 & 1 & 8 & 0.33 & 0.33 & 97.42\% & 2678.60  & 3 & 1 & 8 & 0.33 & 0.33 & 91.95\% & 2655.95  \\
2 & 2 & 0 & 1 & 1 & 92.99\% & 485.425  & 3 & 2 & 0 & 1 & 1 & 94.61\% & 545.25   \\
2 & 2 & 0 & 1 & 0.5 & 93.45\% & 448.193  & 3 & 2 & 0 & 1 & 0.5 & 95.00\% & 564.94   \\
2 & 2 & 0 & 1 & 0.33 & 94.71\% & 412.517  & 3 & 2 & 0 & 1 & 0.33 & 94.31\% & 577.94   \\
2 & 2 & 0 & 0.5 & 1 & 91.11\% & 449.145  & 3 & 2 & 0 & 0.5 & 1 & 95.45\% & 649.61   \\
2 & 2 & 0 & 0.5 & 0.5 & 93.87\% & 426.688  & 3 & 2 & 0 & 0.5 & 0.5 & 97.14\% & 532.09   \\
2 & 2 & 0 & 0.5 & 0.33 & 95.50\% & 389.202  & 3 & 2 & 0 & 0.5 & 0.33 & 94.11\% & 572.81   \\
2 & 2 & 0 & 0.33 & 1 & 91.87\% & 441.520  & 3 & 2 & 0 & 0.33 & 1 & 95.99\% & 585.26   \\
2 & 2 & 0 & 0.33 & 0.5 & 95.03\% & 432.876  & 3 & 2 & 0 & 0.33 & 0.5 & 96.41\% & 519.03   \\
2 & 2 & 0 & 0.33 & 0.33 & 95.74\% & 353.907  & 3 & 2 & 0 & 0.33 & 0.33 & 96.68\% & 535.36   \\
2 & 2 & 2 & 1 & 1 & 90.59\% & 772.53   & 3 & 2 & 2 & 1 & 1 & 90.59\% & 882.89   \\
2 & 2 & 2 & 1 & 0.5 & 93.20\% & 648.75   & 3 & 2 & 2 & 1 & 0.5 & 92.24\% & 923.63   \\
2 & 2 & 2 & 1 & 0.33 & 92.58\% & 634.11   & 3 & 2 & 2 & 1 & 0.33 & 91.25\% & 853.95   \\
2 & 2 & 2 & 0.5 & 1 & 88.43\% & 775.68   & 3 & 2 & 2 & 0.5 & 1 & 93.61\% & 889.52   \\
2 & 2 & 2 & 0.5 & 0.5 & 94.40\% & 677.47   & 3 & 2 & 2 & 0.5 & 0.5 & 94.21\% & 1173.93  \\
2 & 2 & 2 & 0.5 & 0.33 & 94.79\% & 707.98   & 3 & 2 & 2 & 0.5 & 0.33 & 92.16\% & 1081.97  \\
2 & 2 & 2 & 0.33 & 1 & 87.43\% & 715.98   & 3 & 2 & 2 & 0.33 & 1 & 93.13\% & 1100.29  \\
2 & 2 & 2 & 0.33 & 0.5 & 94.09\% & 650.83   & 3 & 2 & 2 & 0.33 & 0.5 & 95.76\% & 919.08   \\
2 & 2 & 2 & 0.33 & 0.33 & 93.18\% & 602.58   & 3 & 2 & 2 & 0.33 & 0.33 & 92.40\% & 787.88   \\
2 & 2 & 6 & 1 & 1 & 85.26\% & 2106.61  & 3 & 2 & 6 & 1 & 1 & 87.81\% & 1673.69  \\
2 & 2 & 6 & 1 & 0.5 & 90.67\% & 2081.86  & 3 & 2 & 6 & 1 & 0.5 & 91.83\% & 2500.54  \\
2 & 2 & 6 & 1 & 0.33 & 93.43\% & 2286.87  & 3 & 2 & 6 & 1 & 0.33 & 93.82\% & 1517.69  \\
2 & 2 & 6 & 0.5 & 1 & 85.20\% & 1992.59  & 3 & 2 & 6 & 0.5 & 1 & 89.60\% & 1801.03  \\
2 & 2 & 6 & 0.5 & 0.5 & 93.06\% & 1706.00  & 3 & 2 & 6 & 0.5 & 0.5 & 92.77\% & 1627.68  \\
2 & 2 & 6 & 0.5 & 0.33 & 93.67\% & 2178.80  & 3 & 2 & 6 & 0.5 & 0.33 & 93.05\% & 1571.33  \\
2 & 2 & 6 & 0.33 & 1 & 82.58\% & 2804.54  & 3 & 2 & 6 & 0.33 & 1 & 88.30\% & 1723.68  \\
2 & 2 & 6 & 0.33 & 0.5 & 93.10\% & 2286.08  & 3 & 2 & 6 & 0.33 & 0.5 & 94.98\% & 1412.99  \\
2 & 2 & 6 & 0.33 & 0.33 & 94.23\% & 2140.95  & 3 & 2 & 6 & 0.33 & 0.33 & 92.95\% & 1567.71  \\
2 & 2 & 8 & 1 & 1 & 87.29\% & 3582.03  & 3 & 2 & 8 & 1 & 1 & 84.77\% & 2696.18  \\
2 & 2 & 8 & 1 & 0.5 & 90.56\% & 3347.02  & 3 & 2 & 8 & 1 & 0.5 & 90.49\% & 3469.51  \\
2 & 2 & 8 & 1 & 0.33 & 91.34\% & 3235.38  & 3 & 2 & 8 & 1 & 0.33 & 89.59\% & 3762.90  \\
2 & 2 & 8 & 0.5 & 1 & 89.02\% & 3682.26  & 3 & 2 & 8 & 0.5 & 1 & 86.96\% & 2995.31  \\
2 & 2 & 8 & 0.5 & 0.5 & 92.34\% & 3034.45  & 3 & 2 & 8 & 0.5 & 0.5 & 90.24\% & 2699.46  \\
2 & 2 & 8 & 0.5 & 0.33 & 93.06\% & 3116.88  & 3 & 2 & 8 & 0.5 & 0.33 & 90.86\% & 2563.88  \\
2 & 2 & 8 & 0.33 & 1 & 89.28\% & 4602.25  & 3 & 2 & 8 & 0.33 & 1 & 85.99\% & 2809.23  \\
2 & 2 & 8 & 0.33 & 0.5 & 93.11\% & 2949.69  & 3 & 2 & 8 & 0.33 & 0.5 & 93.79\% & 2718.60  \\
2 & 2 & 8 & 0.33 & 0.33 & 92.27\% & 3249.27  & 3 & 2 & 8 & 0.33 & 0.33 & 91.76\% & 2335.32  \\
\hline
\end{longtable}
 
\section{Model Robust Partial Profile Design}\label{sec:robust design}
In this section, we demonstrate how to use our SA algorithm to construct a model-robust Bayesian $\mathcal{D}$-optimal partial profile design and compare its structure with those of the main-effects and interaction-effects designs. 
Subsequently, we evaluate the stability of these three design strategies under under various misspecifications of the model and prior information by evaluating their performance through relative $\mathcal{D}_B$-efficiency.
\subsection{Design Construction}

We consider a DCE consisting of 24 choice sets, each containing two profiles. The experiment includes six attributes, each with levels distributed as 2, 2, 2, 3, 3, 3. In each choice set, one attribute remains constant across all profiles.
Effects-type coding is again employed for all attributes. Consequently, the main-effects parameter dimension $m_{main} = 2+2+2+3+3+3-6=9$.
In the interaction-effects model, in addition to the main effects, we incorporate two-way interactions between the first attribute and the second and fourth attributes, thereby increasing the parameter dimension $m_{int}$ to 12.

Regarding the prior distribution of the main-effects model, we set the prior mean $\boldsymbol{\beta}_{0}^{main}$ as
$$\boldsymbol{\beta}_{0}^{main}=(-1, -1, -1, -1, 0, -1, 0,-1, 0)^{T},$$
and the prior covariance matrix $\boldsymbol{\Sigma}_0^{main}$ as
$$\boldsymbol{\Sigma}_0^{main} = \begin{pmatrix}
1 & 0 & 0 & 0 & 0 & 0 & 0 & 0 & 0 \\
0 & 1 & 0 & 0 & 0 & 0 & 0 & 0 & 0 \\
0 & 0 & 1 & 0 & 0 & 0 & 0 & 0 & 0 \\
0 & 0 & 0 & 1 & -0.5 & 0 & 0 & 0 & 0 \\
0 & 0 & 0 & -0.5 & 1 & 0 & 0 & 0 & 0 \\
0 & 0 & 0 & 0 & 0 & 1 & -0.5 & 0 & 0 \\
0 & 0 & 0 & 0 & 0 & -0.5 & 1 & 0 & 0 \\
0 & 0 & 0 & 0 & 0 & 0 & 0 & 1 & -0.5 \\
0 & 0 & 0 & 0 & 0 & 0 & 0 & -0.5 & 1 \\
\end{pmatrix}.$$
For interaction effects, we again employed a naive prior, setting the prior mean to zero and the variance to one. That is, $\boldsymbol{\beta}_{0}^{int} =(\boldsymbol{\beta}_{0}^{main},\mathbf{0}_{8})$ and $\boldsymbol{\Sigma}_0^{int} =\begin{bmatrix}
   \boldsymbol{\Sigma}_0^{main}&\\
    &{\bf I}_{8}\end{bmatrix}$. 
Based on the prior information discussed above, we applied the three different Bayesian 
$\mathcal{D}$--optimality criteria outlined in Eqs. (\ref{eq:D main}), (\ref{eq:D int}), and (\ref{eq:D robust}) as the objective functions for the SA algorithm to generate the corresponding optimal partial profile designs.
For each design, the SA algorithm employed an adaptive stopping criterion, which terminated the optimization process when no improvement in the $\mathcal{D}_B$ value was observed over an entire reheating cycle.

Table \ref{tab:2profile_1constant} presents the three partial profile designs obtained using different Bayesian $\mathcal{D}$-optimality criteria, where the constant attribute is highlighted in gray.

In the main-effects design, the constant attribute is almost evenly distributed among the three 2-level attributes, with no overlap observed among the 3-level attributes. This finding aligns with \cite{cuervo2016integrated}, as when only main effects are considered, the 3-level attributes require more parameters to be estimated. Consequently, they need to vary more frequently to ensure accurate estimation of their parameter values.

In the interaction-effects design, the first attribute is fixed significantly more often, appearing as the constant attribute in 15 choice sets. Additionally, the 3-level attribute 4 is fixed in 2 choice sets. 
This observation is consistent with the findings of \cite{yu2008model}, which indicate that attributes involved in interaction terms are more likely to exhibit overlap to provide the necessary contrasts for estimating interaction effects \citep{Zwerina1996AGM}.

The robust design exhibits a structure that falls between the two. On one hand, similar to the main-effects design, attribute overlap occurs only among the 2-level attributes. On the other hand, as in the interaction-effects model, the first attribute, which is involved in multiple interaction terms, is fixed more frequently.
Correspondingly, the third attribute, which does not participate in interaction effects, is fixed less often than in the main-effects design.

\begin{center}
\begin{longtable}{ccccccccccccccccccccc}
\caption{Comparison of partial profile designs using three different strategies for DCEs with 24 choice sets, each containing 2 profiles and 1 constant attribute.}
\label{tab:2profile_1constant}\\
    
\toprule
\multirow{2}{*}{Choice set}&  \multicolumn{6}{c}{Main-effects }&  &  \multicolumn{6}{c}{Interaction-effects}& & \multicolumn{6}{c}{Robust }\\
\cline{2-7} \cline{9-14} \cline{16-21}
& 1 &2 & 3 & 4 & 5 & 6 & & 1 & 2 & 3 & 4 & 5 & 6 & & 1 & 2 & 3 & 4 & 5 & 6 \\
\hline
1&\cellcolor{gray}2&  1&  1&  2&  3&  3&  
&\cellcolor{gray}1&  2& 1& 2& 2& 3& 
& 2& 2&\cellcolor{gray}1& 2& 1&1
\\
1&\cellcolor{gray}2&  2&  2&  3&  1&  1&  
&\cellcolor{gray}1&  1& 2& 3& 3& 1& 
& 1& 1&\cellcolor{gray}1& 1& 3&3
\\\hline
2&\cellcolor{gray}1&  2&  1&  1&  2&  2&  
&\cellcolor{gray}2&  1& 1& 3& 1& 3& 
& 2&\cellcolor{gray}2& 1& 3& 1&3
\\
2&\cellcolor{gray}1&  1&  2&  2&  1&  1&  
&\cellcolor{gray}2&  2& 2& 1& 2& 2& 
& 1&\cellcolor{gray}2& 2& 2& 3&2
\\\hline
3&1&\cellcolor{gray}2&  2&  2&  1&  2&  
&2&  1&\cellcolor{gray}2& 3& 3& 3& 
&\cellcolor{gray}2& 1& 2& 3& 2&1
\\
3&  2&\cellcolor{gray}2&  1&  1&  2&  3&  
&  1&  2&\cellcolor{gray}2& 2& 2& 2& 
&\cellcolor{gray}2& 2& 1& 2& 1&3
\\\hline
4&  1&  2&\cellcolor{gray}2&  2&  1&  3&  
&  1&  2&\cellcolor{gray}2& 3& 1& 2& 
&\cellcolor{gray}2& 2& 2& 1& 1&1
\\
4&  2&  1&\cellcolor{gray}2&  1&  2&  2&  
&  2&  1&\cellcolor{gray}2& 1& 2& 1& 
&\cellcolor{gray}2& 1& 1& 3& 2&2
\\\hline
5&  2&\cellcolor{gray}1&  2&  1&  2&  2&  
&\cellcolor{gray}2&  2& 1& 2& 1& 3& 
&\cellcolor{gray}1& 2& 2& 1& 1&2
\\
5&  1&\cellcolor{gray}1&  1&  3&  3&  1&  
&\cellcolor{gray}2&  1& 2& 3& 2& 1& 
&\cellcolor{gray}1& 1& 1& 3& 2&3
\\\hline
6&\cellcolor{gray}1& 1& 1& 3& 3& 2& 
&\cellcolor{gray}2& 2& 1& 2& 3& 1& 
& 1&\cellcolor{gray}1& 1& 3& 3&2
\\
6&\cellcolor{gray}1& 2& 2& 1& 1& 3& 
&\cellcolor{gray}2& 1& 2& 3& 1& 2& 
& 2&\cellcolor{gray}1& 2& 2& 1&3
\\\hline
7&\cellcolor{gray}1& 2& 1& 1& 2& 3& 
& 2&\cellcolor{gray}2& 1& 1& 1& 2& 
& 1& 2&\cellcolor{gray}1& 1& 2&1
\\
7&\cellcolor{gray}1& 1& 2& 3& 3& 2& 
& 1&\cellcolor{gray}2& 2& 2& 2& 1& 
& 2& 1&\cellcolor{gray}1& 2& 3&2
\\\hline
8& 1&\cellcolor{gray}1& 2& 1& 1& 3& 
&\cellcolor{gray}2& 1& 2& 2& 1& 3& 
&\cellcolor{gray}2& 1& 1& 2& 3&2
\\
8& 2&\cellcolor{gray}1& 1& 2& 2& 2& 
&\cellcolor{gray}2& 2& 1& 1& 3& 1& 
&\cellcolor{gray}2& 2& 2& 1& 2&3
\\\hline
9&\cellcolor{gray}1& 1& 1& 3& 1& 3& 
&\cellcolor{gray}1& 1& 2& 1& 3& 2& 
& 1& 2&\cellcolor{gray}1& 2& 1&1
\\
9&\cellcolor{gray}1& 2& 2& 1& 2& 1& 
&\cellcolor{gray}1& 2& 1& 3& 1& 1& 
& 2& 1&\cellcolor{gray}1& 1& 2&2
\\\hline
10&\cellcolor{gray}1& 2& 1& 2& 3& 1& 
&\cellcolor{gray}1& 1& 2& 2& 2& 3& 
&\cellcolor{gray}1& 1& 2& 1& 3&2
\\
10&\cellcolor{gray}1& 1& 2& 3& 1& 3& 
&\cellcolor{gray}1& 2& 1& 1& 1& 2& 
&\cellcolor{gray}1& 2& 1& 2& 2&3
\\\hline
11&\cellcolor{gray}2& 1& 2& 1& 2& 1& 
& 2&\cellcolor{gray}2& 1& 3& 1& 2& 
&\cellcolor{gray}2& 2& 1& 2& 3&1
\\
11&\cellcolor{gray}2& 2& 1& 2& 1& 2& 
& 1&\cellcolor{gray}2& 2& 1& 3& 3&
&\cellcolor{gray}2& 1& 2& 1& 1&3
\\\hline
12& 1& 2&\cellcolor{gray}2& 3& 3& 2& 
& 2&\cellcolor{gray}2& 2& 2& 1& 1& 
& 1&\cellcolor{gray}2& 2& 2& 1&2
\\
12& 2& 1&\cellcolor{gray}2& 2& 2& 1& 
& 1&\cellcolor{gray}2& 1& 3& 3& 2& 
& 2&\cellcolor{gray}2& 1& 1& 3&1
\\\hline
13& 1&\cellcolor{gray}2& 2& 2& 2& 1& 
&\cellcolor{gray}1& 2& 2& 3& 1& 1& 
& 1&\cellcolor{gray}2& 2& 1& 3&3
\\
13& 2&\cellcolor{gray}2& 1& 1& 1& 2& 
&\cellcolor{gray}1& 1& 1& 1& 3& 3& 
& 2&\cellcolor{gray}2& 1& 3& 1&2
\\\hline
14& 2&\cellcolor{gray}2& 1& 3& 1& 1& 
& 1& 1& 2&\cellcolor{gray}1& 3& 1& 
& 1&\cellcolor{gray}1& 2& 3& 1&3
\\
14& 1&\cellcolor{gray}2& 2& 1& 3& 3& 
& 2& 2& 1&\cellcolor{gray}1& 1& 3& 
& 2&\cellcolor{gray}1& 1& 1& 2&1
\\\hline
15& 1& 2&\cellcolor{gray}2& 2& 2& 1& 
&\cellcolor{gray}1& 2& 2& 2& 1& 2& 
&\cellcolor{gray}1& 2& 2& 3& 1&1
\\
15& 2& 1&\cellcolor{gray}2& 1& 3& 3& 
&\cellcolor{gray}1& 1& 1& 3& 2& 3& 
&\cellcolor{gray}1& 1& 1& 2& 3&3
\\\hline
16& 1& 1&\cellcolor{gray}1& 2& 3& 2& 
& 1& 2&\cellcolor{gray}2& 1& 1& 3& 
&\cellcolor{gray}1& 2& 1& 1& 3&1
\\
16& 2& 2&\cellcolor{gray}1& 3& 1& 1& 
& 2& 1&\cellcolor{gray}2& 2& 2& 1& 
&\cellcolor{gray}1& 1& 2& 3& 2&2
\\\hline
17&\cellcolor{gray}2& 1& 2& 3& 2& 2& 
&\cellcolor{gray}2& 2& 2& 1& 1& 1& 
& 1&\cellcolor{gray}2& 1& 3& 1&3
\\
17&\cellcolor{gray}2& 2& 1& 1& 3& 3& 
&\cellcolor{gray}2& 1& 1& 2& 3& 2& 
& 2&\cellcolor{gray}2& 2& 1& 2&1
\\\hline
18& 1& 2&\cellcolor{gray}2& 1& 3& 2& 
&\cellcolor{gray}1& 1& 2& 2& 1& 1& 
&\cellcolor{gray}1& 2& 1& 1& 2&2
\\
18& 2& 1&\cellcolor{gray}2& 2& 2& 3& 
&\cellcolor{gray}1& 2& 1& 1& 2& 2& 
&\cellcolor{gray}1& 1& 2& 2& 1&1
\\\hline
19&\cellcolor{gray}2& 2& 1& 3& 3& 1& 
& 1& 1& 2&\cellcolor{gray}3& 2& 2& 
& 2& 1&\cellcolor{gray}2& 2& 1&2
\\
19&\cellcolor{gray}2& 1& 2& 1& 1& 2& 
& 2& 2& 1&\cellcolor{gray}3& 3& 1& 
& 1& 2&\cellcolor{gray}2& 3& 3&1
\\\hline
20& 2& 1&\cellcolor{gray}2& 1& 3& 1& 
&\cellcolor{gray}1& 2& 2& 1& 1& 1& 
& 1&\cellcolor{gray}1& 2& 1& 2&3
\\
20& 1& 2&\cellcolor{gray}2& 2& 2& 2& 
&\cellcolor{gray}1& 1& 1& 2& 3& 2& 
& 2&\cellcolor{gray}1& 1& 3& 3&1
\\\hline
21& 1& 2&\cellcolor{gray}1& 1& 1& 2& 
& 1&\cellcolor{gray}1& 2& 2& 1& 2& 
&\cellcolor{gray}2& 2& 1& 3& 3&2
\\
21& 2& 1&\cellcolor{gray}1& 2& 3& 1& 
& 2&\cellcolor{gray}1& 1& 1& 2& 1& 
&\cellcolor{gray}2& 1& 2& 2& 2&1
\\\hline
22& 1&\cellcolor{gray}2& 2& 1& 3& 2& 
&\cellcolor{gray}1& 1& 2& 1& 2& 3& 
& 2&\cellcolor{gray}1& 1& 1& 1&3
\\
22& 2&\cellcolor{gray}2& 1& 2& 1& 3& 
&\cellcolor{gray}1& 2& 1& 2& 3& 1& 
& 1&\cellcolor{gray}1& 2& 2& 2&1
\\\hline
23& 2& 2&\cellcolor{gray}2& 1& 1& 1& 
&\cellcolor{gray}2& 2& 1& 3& 2& 2& 
&\cellcolor{gray}1& 1& 2& 3& 1&1
\\
23& 1& 1&\cellcolor{gray}2& 3& 2& 3& 
&\cellcolor{gray}2& 1& 2& 1& 1& 3& 
&\cellcolor{gray}1& 2& 1& 2& 2&2
\\\hline
24& 2&\cellcolor{gray}2& 2& 2& 1& 2& 
&\cellcolor{gray}2& 2& 1& 3& 1& 1& 
&\cellcolor{gray}2& 1& 2& 2& 2&1
\\
24& 1&\cellcolor{gray}2& 1& 3& 2& 3& 
&\cellcolor{gray}2& 1& 2& 1& 3& 2& 
&\cellcolor{gray}2& 2& 1& 1& 1&2
\\
\hline
\end{longtable}
\end{center}
\subsection{Sensitivity to Model and Prior Misspecifications}
We compare the performance of the three designs presented in Table \ref{tab:2profile_1constant} under the following four true model specifications:
\begin{enumerate}
\item Model I: No interaction terms are present, meaning the true prior mean is given by 
$$\boldsymbol{\beta}_0^{true} = (-1, -1, -1, -1, 0, -1, 0,-1, 0)^{T}.$$
\item Model II: Only the interaction between Attribute 1 and Attribute 2 is present, with the true prior mean given by 
$$\boldsymbol{\beta}_0^{true} = (-1, -1, -1, -1, 0, -1, 0,-1, 0, \lambda)^{T}.$$
\item Model III: Only the interaction between Attribute 1 and Attribute 4 is present, with the true prior mean given by 
$$\boldsymbol{\beta}_0^{true} = (-1, -1, -1, -1, 0, -1, 0,-1, 0, \lambda, \lambda)^{T}.$$
\item Model IV: Interactions between Attribute 1 and both Attributes 2 and 4 are present, with the true prior mean given by 
$$\boldsymbol{\beta}_0^{true} = (-1, -1, -1, -1, 0, -1, 0,-1, 0, \lambda, \lambda, \lambda)^{T}.$$
\end{enumerate}
In this setup, $\lambda$ acts as a scale factor for the interaction effects, with $\lambda \in \{0.1, 0.2, 0.3\}$. For each true model specification, we vary the value of $\lambda$, resulting in a total of 10 different true prior means.

Regarding the true prior variance, the prior covariance matrix for the main effects is set as $\boldsymbol{\Sigma}_0^{main}$. For the interaction effects, we assume a prior variance of zero to precisely assess the impact of the interaction size on the design efficiency.

Using each true prior distribution, we generate a benchmark partial profile design that serves as a reference under the corresponding prior specification.
For each case, we compare the $\mathcal{D}_{B}$-efficiency of the three designs from Table \ref{tab:2profile_1constant} relative to the corresponding benchmark design. Notably, under Model I, the benchmark design is set as the main-effects design, implying that in this scenario, the relative $\mathcal{D}_{B}$-efficiency of the main-effects design is equal to 1.

\begin{table}[h]
\caption{Comparison of $\mathcal{D}_B$-efficiency of the three design strategies relative to the benchmark design in different DCE settings.}
    \label{tab:DB-Eff_3designs}
    \centering
   \begin{tabular}{ccccc}
\hline
         Model&  $\lambda$&  Main-effect&  Interaction-effect& Robust\\\hline
         I&  -&  100.00\%&  87.23\%& 97.88\%
\\
         II&  0.1&  90.94\%&  89.66\%& 97.79\%
\\
         II&  0.2&  90.85\%&  89.59\%& 97.46\%
\\
         II&  0.3&  90.14\%&  88.96\%& 96.42\%
\\
         III&  0.1&  82.83\%&  93.24\%& 97.04\%
\\
         III&  0.2&  81.57\%&  91.94\%& 96.25\%
\\
         III&  0.3&  80.41\%&  90.84\%& 95.63\%
\\
         IV&  0.1&  76.01\%&  92.32\%& 95.03\%
\\
         IV&  0.2&  75.56\%&  91.63\%& 95.02\%
\\
 IV& 0.3& 74.33\%& 89.90\%&93.93\%
\\\hline
 \multicolumn{2}{c}{Average}& 84.26\%& 90.53\%&96.25\%\\\hline
    \end{tabular}
    
\end{table}

Table \ref{tab:DB-Eff_3designs} presents the relative $\mathcal{D}_B$-efficiency of the three designs compared to the benchmark design across 10 different scenarios. 
The results show that the main-effects design performs worse when interaction effects are present in the underlying model. Specifically, as the number and magnitude of interaction effects increase, the efficiency of the main-effects design declines further.
Conversely, the interaction-effects design exhibits the worst relative performance when the number of interaction terms is small. However, as the number of interaction terms increases, its relative performance improves.

The robust design consistently demonstrates strong performance across all considered scenarios, with an average relative $\mathcal{D}_B$-efficiency of 96.25\%. Notably, when interaction effects are present, its relative $\mathcal{D}_B$-efficiency is the highest among the three designs in every case. 

These findings highlight that the robust design is significantly more stable against model misspecification than the main-effects and interaction-effects designs.
When prior information is limited and the underlying model is uncertain, employing the model-robust Bayesian $\mathcal{D}$-optimality criterion for constructing partial profile designs proves to be a more reliable approach.

\section{Healthcare Intervention Case Study}\label{sec:real-life}

In this section, we revisit the healthcare intervention DCE conducted by \cite{Luyten2015}, identifying potential experimental design issues outlined in Section \ref{Sec:example}. To address these concerns, we propose several alternative partial profile designs and compare their Bayesian $\mathcal{D}$-efficiency. Subsequently, we conduct simulation experiments to evaluate the accuracy of parameter estimation across these designs.
\subsection{Case Study Setup}\label{subsec: case study setup}
At the experimental design stage, \cite{Luyten2015} constructed a Bayesian $\mathcal{D}$-optimal design based on an interaction-effects MNL model which includes 27 parameters. 
The prior mean for the main effects was specified as $\boldsymbol{\beta}_{0}^{main}=(-0.4,-0.5,0,-0.4,0.1,-0.8,0,-0.5,0,-0.5,0.2,-0.5,-0.25,0,0.25)^{T}$. For the prior variance-covariacne matrix, they let $\boldsymbol{\Sigma}_{0}^{main}= \begin{bmatrix}
A_1 &&&&&&\\
&A_2&&&&&\\
&&A_2&&&&\\
&&&A_2&&&\\
&&&&A_2&&\\
&&&&&A_2&\\
&&&&&&A_3\\
\end{bmatrix}$, where $A_1=[0.09]$, $A_2=\begin{bmatrix}
    0.09&-0.045\\
    -0.045&0.09\end{bmatrix}$, and $A_3 = \begin{bmatrix} 
    0.09&-0.0225&-0.0225&-0.0225\\
    -0.0225&0.09&-0.0225&-0.0225\\
    -0.0225&-0.0225&0.09&-0.0225\\
    -0.0225&-0.0225&-0.0225&0.09
    \end{bmatrix}$. Regarding the interaction effects, they assumed both the prior mean and variance to be 0. Using this prior information, they constructed a Bayesian optimal design for the 42 choice sets. 

After combining the data from the three survey groups, they initially analyzed it using the same 27-parameter MNL model employed during the experimental design stage. 
Using likelihood ratio tests, they found that only the interaction terms between $x_1$ and $x_4$ as well as $x_1$ and $x_7$ were statistically significant.  
After dropping the non-significant interaction terms, they obtained a final model with the number of parameters reduced from 27 to 21 compared to the model used in the experimental design. 
Our case study evaluates the performance of different partial profile designs when lacking prior information for the interaction effects. 
\subsection{Alternative Partial Profile Designs}
We consider four different Bayesian $\mathcal{D}$-optimal partial profile designs:
\begin{enumerate}
\item Main-effects design: This is a Bayesian partial profile design that focuses solely on the main effects of the attributes. To calculate the Bayesian $\mathcal{D}$-optimality criterion, we used the prior distribution $\mathrm{N}(\boldsymbol{\beta}_{0}^{main},\boldsymbol{\Sigma}_0^{main})$, where the specific values of $\boldsymbol{\beta}_{0}^{main}$ and $\boldsymbol{\Sigma}_0^{main}$ are provided earlier in Section \ref{subsec: case study setup}.
\item Original design: This is the original design employed in \cite{Luyten2015}, which is detailed in Table \ref{tab:health care org designs}.
\item Robust design: This is a Bayesian partial profile design that seeks to minimize the composite design criterion defined in Eq. (\ref{eq:D robust}). For the main-effects part, we also used the same prior parameters $\boldsymbol{\beta}_{0}^{main}$ and $\boldsymbol{\Sigma}_0^{main}$. As for the interaction-effects part, we also consider all the two-way interactions of interest and apply a naive prior distribution, where the prior mean is set to zero, and the variance is set to one. That is, $\boldsymbol{\beta}_{\boldsymbol{0}}^{int} =(\boldsymbol{\beta}_{0}^{main},\mathbf{0}_{12})$ and $\boldsymbol{\Sigma}_0^{int} =\begin{bmatrix}
    \boldsymbol{\Sigma}_0^{main}&\\
    &{\bf I}_{12}\end{bmatrix}$.
\item True model design: A benchmark design was generated based on the final model obtained after the likelihood ratio tests, incorporating all main effects and only the interactions between $x_1$ and $x_4$ as well as $x_1$ and $x_7$, representing an ideal scenario where information on interaction effects is fully available. 
For consistency, we used the same prior  $\boldsymbol{\beta}_{0}^{main}$ and $\boldsymbol{\Sigma}_0^{main}$ for main effects. Regarding interaction effects, we used the results of the data analysis from the study of \cite{Luyten2015} as prior information. The specific values are available in Table \ref{tab:true_prior}. Combining the two, we obtained the prior distribution for the true model, denoted as $\mathrm{N}(\boldsymbol{\beta}_{0}^{true},\boldsymbol{\Sigma}_0^{true})$.
\begin{table}[h]

    \centering
     \caption{Prior information of the interaction effects in the true model.}
    \begin{tabular}{lcccccc}
    \hline
          &$x_{11}x_{41}$&  $x_{11}x_{42}$&  $x_{11}x_{71}$&  $x_{11}x_{72}$&  $x_{11}x_{73}$& $x_{11}x_{74}$\\\hline
          Mean&-0.0431&  0.0345&  0.012&  -0.0676&  -0.048& 0.1103\\
          SD &0.0378&  0.0394&  0.0528&  0.0524&  0.0558& 0.0578\\
          \hline
    \end{tabular}
   
    \label{tab:true_prior}
\end{table}
\end{enumerate}
Using the prior information mentioned above, we employed our SA algorithm to construct the main-effects design, robust design, and true model design. 
Similar to the work of \cite{Luyten2015}, we also excluded unrealistic attribute level combinations. The specific details of these designs are detailed in Table \ref{tab:health care designs} in Appendix B.
\begin{table}[h]
    \centering
\caption{$\mathcal{D}_B$-efficiency of four different designs relative to the true model design.}
\label{tab:health care DB}
    \begin{tabular}{lcccc}
    \hline
&  Main-effects&  Original&  Robust&  True model\\
\hline
$\mathcal{D}_B$-efficiency&  75.89\%&  78.64\%&  89.70\%& 100\%\\
 \hline
\end{tabular}

\end{table}

Table \ref{tab:health care DB} records the Bayesian $\mathcal{D}_B$-efficiency of each design relative to the true model design.
The main-effects design performed the worst, with an efficiency of only 75.89\% relative to the true model design. This emphasizes the importance of considering interaction effects during the experimental design stage. 
The original design also performed poorly, only slightly better than the main-effects design.
This partly due to the limitations of the design algorithm and partly to the inclusion of an excessive number of interaction terms not present in the true model. 
In contrast, the robust design performed better, achieving a $\mathcal{D}_B$-efficiency of nearly 90\% relative to the true model design.

\subsection{Comparison of Parameter Estimate Accuracy}
We evaluated the estimation accuracy of different designs through simulation experiments. Specifically, we assumed a total of 300 participants in the DCE, evenly distributed across three survey groups.
For each partial profile design, we simulated responses under the assumption that $\boldsymbol{\beta}_{0}^{true}$ represents the true parameter vector, denoted as $\boldsymbol{\beta^{*}}$. 
The choice data for each design were then analyzed using an MNL model, and the corresponding parameter estimates were recorded.
To ensure the robustness of our results, we conducted 500 independent simulations, generating 500 datasets and the associated parameter estimates for each design.

To compare the accuracy of parameter estimates across different designs, we use the expected mean square error (EMSE) as a measure, which can be defined as 
\begin{equation}\label{eq:EMSEbeta}
    \text{EMSE}_{\hat{\boldsymbol{\beta}}}(\boldsymbol{\beta^{*}}) =  
    \int_{\mathcal{R}^{k}} 
    \left(\hat{\boldsymbol{\beta}} - \boldsymbol{\beta^{*}}\right)^{T}
    \left(\hat{\boldsymbol{\beta}} - \boldsymbol{\beta^{*}}\right)
    \pi(\hat{\boldsymbol{\beta}}) \, \mathrm{d}\hat{\boldsymbol{\beta}},
\end{equation}
where $\pi(\hat{\boldsymbol{\beta}})$ represents the distribution of the estimates. A smaller $\text{EMSE}_{\hat{\boldsymbol{\beta}}}(\boldsymbol{\beta^{*}})$ value indicates greater accuracy of the estimated parameters. 
As in \cite{kessels2006comparison}, the EMSE value is approximated by
\begin{equation}\label{eq:EMSEbeta2}
    \text{EMSE}_{\hat{\boldsymbol{\beta}}}(\boldsymbol{\beta^{*}}) =  \frac{1}{N} \sum_{n=1}^{N} 
    \left(\hat{\boldsymbol{\beta}}^{n} - \boldsymbol{\beta^{*}}\right)^{T}
    \left(\hat{\boldsymbol{\beta}}^{n} - \boldsymbol{\beta^{*}}\right),
\end{equation}
where $N=500$ is the number of simulation dataset, and $\hat{\boldsymbol{\beta}}^{n}$ denotes the vector of estimates obtained from the $n$-th simulated dataset.

\begin{figure}[h]
\centering
\includegraphics[width=0.8\textwidth]{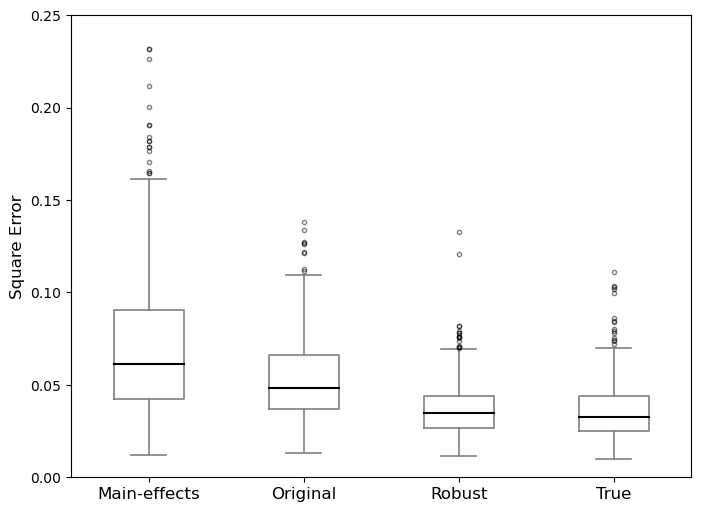}
\caption{Square error of the estimates obtained from different partial profile designs in each simulation.}
\label{fig:boxplot}
\end{figure}

Figure \ref{fig:boxplot} is a boxplot presenting the square error between the parameter estimates and the true values obtained from the 500 simulation experiments for the four different designs.
The black line within each box indicates the median square error for each design.
It is clear from the boxplot that the main design and original design have larger medians and variances compared to the robust design and true model design. The robust design and true model design have similar performance, with the true model design having a slightly lower median. 
\begin{table}[h]
    \centering
\caption{EMSE values comparing estimation accuracy of four different designs.}
\label{tab:health care EMSE}
    \begin{tabular}{lcccc}
    \hline
&  Main-effects&  Original&  Robust&  True model\\
\hline
EMSE &  0.0714&  0.0528&  0.0370& 0.0361\\
 \hline
\end{tabular}

\end{table}

The specific EMSE values for the four designs are presented in Table \ref{tab:health care EMSE}. 
The EMSE of the main design is nearly twice that of the true model design, while the original design's EMSE increases by about 50\%. This indicates that the accuracy of parameter estimates is significantly reduced for these two designs in the face of model misspecification. 
In contrast, the EMSE of the robust design is close to that of the true model design, further validating that SA model-robust design performs relatively well across different underlying models.

\section{Conclusion and Discussion}\label{sec:conclusion}
This paper introduces an integrated SA algorithm for constructing Bayesian optimal partial profile interaction-effects designs. Our SA algorithm starts with an initial random partial profile design. In each iteration, it randomly selects an attribute from a random profile within a choice set, then determines whether the attribute is constant, and generates corresponding random perturbations. Our exploration rule ensures that each new design generated during each iteration satisfies the constraint that each choice set in the partial profile design contains a certain number of constant attributes. 
Compared to other algorithms focused on partial profile interaction-effects designs, our algorithm can handle any number of profiles and does not require the unrealistic assumption that respondents have no specific preference among the profiles. 
Compared to the commonly used two-stage CE algorithm, our SA algorithm takes into account the impact of interaction effects on the distribution of constant attributes and can flexibly optimize both the distribution of constant attributes and attribute levels simultaneously. 
Through extensive computational experiments, we have demonstrated that our SA algorithm consistently outperforms the CE algorithm in generating more efficient designs across all DCE scenarios that we have considered. This advantage is particularly pronounced when the prior variance is large and when the model includes a greater number of interaction terms.

Since information about interaction effects is often limited during the experimental design stage, we suggest a model-robust design strategy to achieve good parameter estimate accuracy regardless of the underlying model. Our robust design employs a composite optimality criterion that accounts for both the main-effects model and the interaction-effects model. Through simulation experiments and a real-life healthcare case study, we demonstrate that our robust design performs well under various cases of model misspecification.

From a practical perspective, incorporating interaction effects into the model significantly increases the number of parameters, while the attribute overlap in partial profile designs generally reduces statistical efficiency. Consequently, obtaining estimates with the same level of precision as a full-profile design often requires a greater number of choice sets.
In such cases, rather than relying on a single design, we recommend generating a larger set of choice sets and distributing them across multiple survey groups. This approach allows for the collection of more statistical information while preventing individual respondents from being overwhelmed by an excessive number of choice tasks.

Future research could extend our work by developing an adaptive partial profile design approach. In most applications of partial profile designs, the number of constant attributes within each choice set remains fixed. However, in practice, as respondents progress through the experiment, they may experience fatigue and boredom, which can in turn introduce order effects and reduce attentiveness \citep{Savage2008Learning,mao2025order}. In such cases, the likelihood of respondents resorting to non-compensatory decision rules may significantly increase.
To deal with this, future studies could explore adaptive partial profile designs where the number of constant attributes varies based on the position of the choice set within the experiment. Specifically, in the earlier choice sets, where respondents' attention is relatively high, fewer attributes could be held constant. Conversely, in later choice sets, where respondent fatigue is more pronounced, increasing the number of fixed attributes could help alleviate cognitive burden and improve response quality.

\clearpage
\bibliography{ref}
\clearpage

\appendix
\renewcommand{\thesection}{Appendix \Alph{section}.}
\setcounter{table}{0}
\setcounter{figure}{0}
\section{Detailed Attribute Levels and Experimental Design of the Healthcare Intervention Discrete Choice Experiment}\label{sec:appA}
\renewcommand{\thetable}{A\arabic{table}}
\renewcommand{\thefigure}{A\arabic{figure}}
\begin{table}[h!]
\caption{Attributes and levels for the healthcare intervention DCE.} \label{tab:health_attributes}
\centering
\begin{tabularx}{\textwidth}{XX}
\hline
Attribute &  Level \\
\hline
$x_1$: What type of intervention is
it? & 1. Preventive (aiming to prevent
healthy persons from\\
& \hspace{1.25em}becoming ill)\\
&2. Curative (aiming to cure
people who are ill)\\
$x_2$: How big is the probability of
success of the  & 1. 1 in 3 is successful (33\%)\\
\hspace{1.75em}intervention?& 2. 2 in 3 is successful (66\%)\\
&3. Always successful (100\%)\\
$x_3$: How often do adverse effects occur? &1. Often\\
&2. Rarely\\
&3. Never\\
$x_4$: How severe is the illness for which the intervention  &1. Not lethal, but everyone who
gets the disease will\\
\hspace{1.75em}is
developed?& \hspace{1.25em}experience a short period of
illness without lasting\\
&\hspace{1.25em}effects (not severe)\\
&2. Not lethal, but everyone who
gets the disease will\\
&\hspace{1.25em}experience a severe and
lasting reduction in quality \\
&\hspace{1.25em}of life (severe)\\
&3. Lethal, everyone who gets the
disease will die from \\
&\hspace{1.25em}it (lethal)\\
$x_5$: Does the patient cause the
disease through his or&1. Fully\\
\hspace{1.75em}her own lifestyle?&2. Partly\\
& 3. Not at all\\
$x_6$: How long does it take before
the patient becomes &1. After 20 years\\
\hspace{1.75em}ill/shows signs/symptoms of illness?&2. After 5 years\\
&3. Within a year\\
$x_7$: At what age does the patient become ill? & 1. 80–90 years\\
&2. 60–70 years\\
&3. 40–50 years\\
&4. 20–30 years\\
&5. 0–10 years\\
\hline
\end{tabularx}
\end{table}

\begin{longtable}{cccccccccccccccccccccccc}
\caption{The original partial profile designs used in the healthcare intervention DCE.}
\label{tab:health care org designs}

      \\
    \hline
        \multirow{2}{*}{Choice set}&  \multicolumn{7}{c}{Survey 1 }&  &  \multicolumn{7}{c}{Survey 2}& & \multicolumn{7}{c}{Survey 3 }\\
         \cline{2-8} \cline{10-16} \cline{18-24}
            &1 & 2 & 3 & 4 & 5 & 6 &7& & 1 & 2 & 3 & 4 & 5 & 6 &7& & 1 & 2 & 3 & 4 & 5 & 6&7 \\
         \hline
                  1
&  \cellcolor{gray}2&  \cellcolor{gray}3&  1&  2&  \cellcolor{gray}1&  3&  5
&  &  \cellcolor{gray}2& \cellcolor{gray}3& \cellcolor{gray}2& 2& 3& 2& 3
& & \cellcolor{gray}2& \cellcolor{gray}3& \cellcolor{gray}2& 1& 1& 3&4
\\
         1
&  \cellcolor{gray}2&  \cellcolor{gray}3&  2&  1&  \cellcolor{gray}1&  2&  3
&  &  \cellcolor{gray}2& \cellcolor{gray}3& \cellcolor{gray}2& 3& 2& 1& 4
& & \cellcolor{gray}2& \cellcolor{gray}3& \cellcolor{gray}2& 3& 2& 2&2
\\\hline
         2
&  \cellcolor{gray}2&  \cellcolor{gray}3&  3&  2&  1&  2&  \cellcolor{gray}1
&  &  \cellcolor{gray}2& \cellcolor{gray}3& 3& 2& 2& 3& \cellcolor{gray}1
& & \cellcolor{gray}2& \cellcolor{gray}3& 3& 3& 1& 3&\cellcolor{gray}2
\\
         2
&  \cellcolor{gray}2&  \cellcolor{gray}3&  1&  1&  3&  3&  \cellcolor{gray}1
&  &  \cellcolor{gray}2& \cellcolor{gray}3& 2& 3& 3& 1& \cellcolor{gray}1
& & \cellcolor{gray}2& \cellcolor{gray}3& 2& 2& 2& 2&\cellcolor{gray}2
\\\hline
         3
&  \cellcolor{gray}1&  2&  \cellcolor{gray}3&  \cellcolor{gray}3&  3&  1&  3
&  &  \cellcolor{gray}1& 3& \cellcolor{gray}2& 1& \cellcolor{gray}3& 3& 1
& & \cellcolor{gray}1& 1& \cellcolor{gray}1& \cellcolor{gray}1& 2& 2&1
\\
         3
&  \cellcolor{gray}1&  3&  \cellcolor{gray}3&  \cellcolor{gray}3&  2&  3&  2
&  &  \cellcolor{gray}1& 2& \cellcolor{gray}2& 2& \cellcolor{gray}3& 1& 2
& & \cellcolor{gray}1& 3& \cellcolor{gray}1& \cellcolor{gray}1& 1& 1&2
\\\hline
         4
&  \cellcolor{gray}1&  2&  \cellcolor{gray}2&  1&  1&  \cellcolor{gray}1&  4
&  &  \cellcolor{gray}1& 2& 2& \cellcolor{gray}1& 1& \cellcolor{gray}3& 5
& & \cellcolor{gray}1& 3& 2& \cellcolor{gray}3& \cellcolor{gray}1& 1&1
\\
         4
&  \cellcolor{gray}1&  1&  \cellcolor{gray}2&  2&  2&  \cellcolor{gray}1&  2
&  &  \cellcolor{gray}1& 1& 1& \cellcolor{gray}1& 3& \cellcolor{gray}3& 4
& & \cellcolor{gray}1& 2& 1& \cellcolor{gray}3& \cellcolor{gray}1& 2&3
\\\hline
         5
&  \cellcolor{gray}1&  3&  2&  \cellcolor{gray}2&  \cellcolor{gray}2&  2&  3
&  &  \cellcolor{gray}1& 1& 2& \cellcolor{gray}2& 3& 2& \cellcolor{gray}1
& & \cellcolor{gray}1& 2& 3& 1& \cellcolor{gray}3& \cellcolor{gray}3&2
\\
         5
&  \cellcolor{gray}1&  1&  1&  \cellcolor{gray}2&  \cellcolor{gray}2&  3&  5
&  &  \cellcolor{gray}1& 2& 3& \cellcolor{gray}2& 2& 1& \cellcolor{gray}1
& & \cellcolor{gray}1& 3& 1& 3& \cellcolor{gray}3& \cellcolor{gray}3&5
\\\hline
 6
& \cellcolor{gray}1& 1& 3& 3& 1& \cellcolor{gray}3& \cellcolor{gray}3
& & \cellcolor{gray}1& 1& 2& 1& \cellcolor{gray}1& \cellcolor{gray}2& 4
& & \cellcolor{gray}1& 1& 3& 3& 2& \cellcolor{gray}1&\cellcolor{gray}4
\\
 6
& \cellcolor{gray}1& 2& 1& 1& 2& \cellcolor{gray}3& \cellcolor{gray}3
& & \cellcolor{gray}1& 2& 3& 2& \cellcolor{gray}1& \cellcolor{gray}2& 1
& & \cellcolor{gray}1& 3& 1& 2& 1& \cellcolor{gray}1&\cellcolor{gray}4
\\\hline
 7
& 2& \cellcolor{gray}3& \cellcolor{gray}1& \cellcolor{gray}3& 1& 2& 2
& & 1& \cellcolor{gray}3& \cellcolor{gray}1& 2& 3& \cellcolor{gray}3& 5
& & 1& \cellcolor{gray}3& \cellcolor{gray}2& 2& 2& \cellcolor{gray}2&3
\\
 7
& 1& \cellcolor{gray}3& \cellcolor{gray}1& \cellcolor{gray}3& 2& 3& 3
& & 2& \cellcolor{gray}3& \cellcolor{gray}1& 3& 2& \cellcolor{gray}3& 3
& & 2& \cellcolor{gray}3& \cellcolor{gray}2& 3& 1& \cellcolor{gray}2&4
\\\hline
 8
& 1& \cellcolor{gray}3& \cellcolor{gray}3& 2& 2& \cellcolor{gray}1& 4
& & 2& \cellcolor{gray}3& 3& \cellcolor{gray}2& \cellcolor{gray}3& 1& 1
& & 2& \cellcolor{gray}3& 1& \cellcolor{gray}2& \cellcolor{gray}1& 1&3
\\
 8
& 2& \cellcolor{gray}3& \cellcolor{gray}3& 1& 3& \cellcolor{gray}1& 2
& & 1& \cellcolor{gray}3& 2& \cellcolor{gray}2& \cellcolor{gray}3& 3& 3
& & 1& \cellcolor{gray}3& 3& \cellcolor{gray}2& \cellcolor{gray}1& 2&4
\\\hline
 9
& 2& \cellcolor{gray}3& 1& \cellcolor{gray}1& 2& 2& \cellcolor{gray}2
& & 2& \cellcolor{gray}3& 2& \cellcolor{gray}1& 2& \cellcolor{gray}3& 5
& & 2& \cellcolor{gray}3& 2& \cellcolor{gray}3& 1& 3&\cellcolor{gray}1
\\
 9
& 1& \cellcolor{gray}3& 3& \cellcolor{gray}1& 3& 1& \cellcolor{gray}2
& & 1& \cellcolor{gray}3& 3& \cellcolor{gray}1& 3& \cellcolor{gray}3& 3
& & 1& \cellcolor{gray}3& 1& \cellcolor{gray}3& 3& 1&\cellcolor{gray}1
\\\hline
 10
& 2& \cellcolor{gray}3& 2& 2& \cellcolor{gray}1& \cellcolor{gray}3& 1
& & 1& \cellcolor{gray}3& 1& 3& \cellcolor{gray}2& 2& \cellcolor{gray}1
& & 1& \cellcolor{gray}3& 3& 1& \cellcolor{gray}2& \cellcolor{gray}3&5
\\
 10
& 1& \cellcolor{gray}3& 3& 1& \cellcolor{gray}1& \cellcolor{gray}3& 5
& & 2& \cellcolor{gray}3& 3& 1& \cellcolor{gray}2& 1& \cellcolor{gray}1
& & 2& \cellcolor{gray}3& 2& 2& \cellcolor{gray}2& \cellcolor{gray}3&4
\\\hline
 11
& 2& 3& \cellcolor{gray}1& 2& \cellcolor{gray}3& 2& \cellcolor{gray}4
& & 1& 2& \cellcolor{gray}2& \cellcolor{gray}2& \cellcolor{gray}1& 3& 1
& & 2& 3& \cellcolor{gray}2& \cellcolor{gray}3& 1& \cellcolor{gray}3&5
\\
 11
& 1& 2& \cellcolor{gray}1& 3& \cellcolor{gray}3& 3& \cellcolor{gray}4
& & 2& 3& \cellcolor{gray}2& \cellcolor{gray}2& \cellcolor{gray}1& 1& 2
& & 1& 2& \cellcolor{gray}2& \cellcolor{gray}3& 2& \cellcolor{gray}3&4
\\\hline
 12
& 1& 2& \cellcolor{gray}2& 3& \cellcolor{gray}3& 2& \cellcolor{gray}2
& & 1& 2& \cellcolor{gray}3& \cellcolor{gray}3& 1& 3& \cellcolor{gray}3
& & 1& 2& \cellcolor{gray}1& 3& \cellcolor{gray}1& 2&\cellcolor{gray}2
\\
 12
& 2& 3& \cellcolor{gray}2& 1& \cellcolor{gray}3& 3& \cellcolor{gray}2
& & 2& 3& \cellcolor{gray}3& \cellcolor{gray}3& 2& 1& \cellcolor{gray}3
& & 2& 3& \cellcolor{gray}1& 2& \cellcolor{gray}1& 1&\cellcolor{gray}2
\\\hline
 13
& 2& 3& 1& \cellcolor{gray}2& \cellcolor{gray}2& \cellcolor{gray}3& 1
& & 1& 1& \cellcolor{gray}3& 2& 3& \cellcolor{gray}2& \cellcolor{gray}4
& & 2& 3& \cellcolor{gray}1& 1& \cellcolor{gray}1& 2&\cellcolor{gray}1
\\
 13
& 1& 2& 2& \cellcolor{gray}2& \cellcolor{gray}2& \cellcolor{gray}3& 4
& & 2& 3& \cellcolor{gray}3& 3& 1& \cellcolor{gray}2& \cellcolor{gray}4
& & 1& 1& \cellcolor{gray}1& 2& \cellcolor{gray}1& 3&\cellcolor{gray}1
\\\hline
 14
& 1& 2& 2& \cellcolor{gray}1& 3& \cellcolor{gray}3& \cellcolor{gray}5
& & 1& 2& 3& 1& \cellcolor{gray}3& \cellcolor{gray}2& \cellcolor{gray}1
& & 1& 1& 2& \cellcolor{gray}1& 2& \cellcolor{gray}1&\cellcolor{gray}3
\\
 14& 2& 3& 3& \cellcolor{gray}1& 1& \cellcolor{gray}3& \cellcolor{gray}5
& & 2& 3& 1& 3& \cellcolor{gray}3& \cellcolor{gray}2& \cellcolor{gray}1
& & 2& 3& 1& \cellcolor{gray}1& 1& \cellcolor{gray}1&\cellcolor{gray}3
\\
\hline

\end{longtable}

\begin{figure}[h!]
\centering
\includegraphics[width=0.85\textwidth]{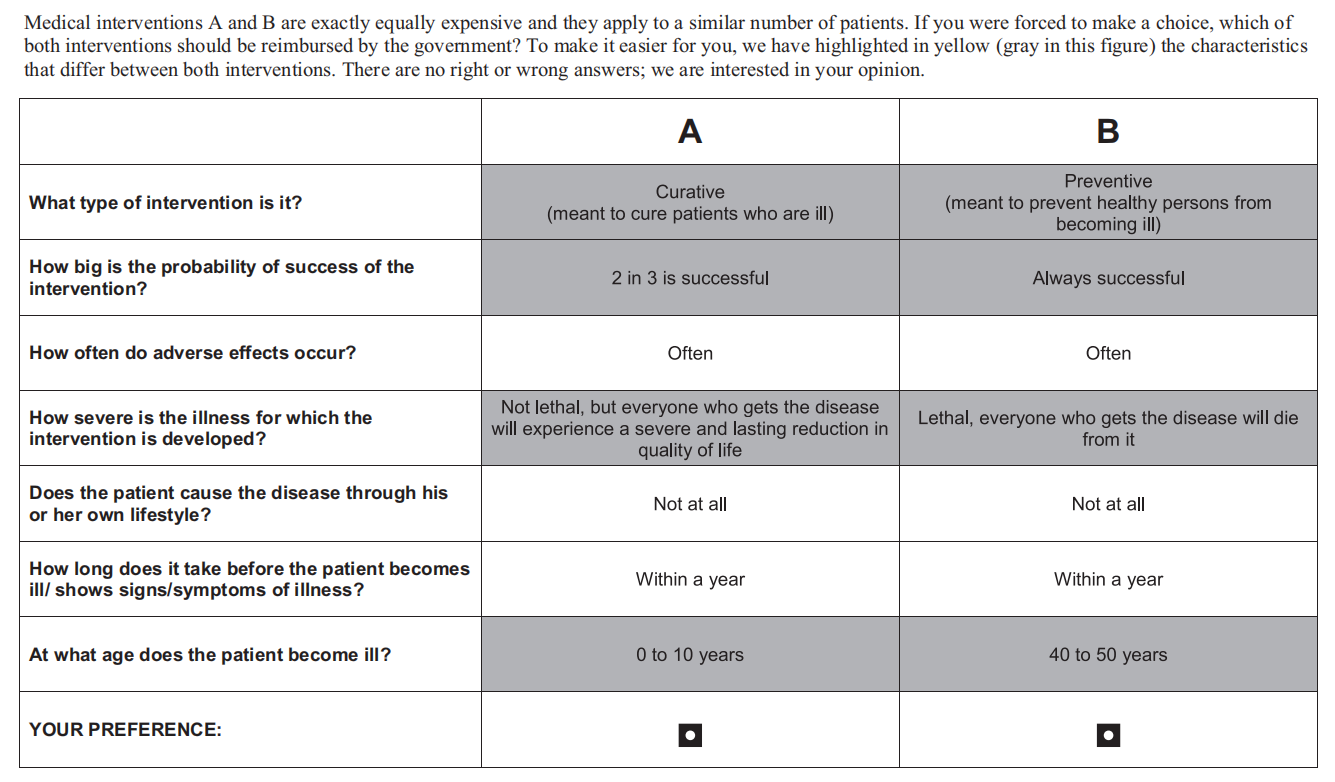}
\caption{Example of a choice set used in the healthcare intervention DCE.}
\label{fig:choice_set}
\end{figure}
\clearpage

\section{Comparison of Partial Profile Designs for the Healthcare Intervention Discrete Choice Experiment.}\label{sec: appB}
\setcounter{table}{0}
\setcounter{figure}{0}
\renewcommand{\thetable}{B\arabic{table}}
\renewcommand{\thefigure}{B\arabic{figure}}

\begin{center}
\begin{longtable}{ccccccccccccccccccccccccc}
\caption{Three different partial profile designs using different strategies for the healthcare intervention DCE.}
\label{tab:health care designs}
      \\
\toprule
\multirow{2}{*}{Survey}& \multirow{2}{*}{Choice set}&  \multicolumn{7}{c}{Main-effects }&  &  \multicolumn{7}{c}{Robust}& & \multicolumn{7}{c}{True model }\\
\cline{3-9} \cline{11-17} \cline{19-25}
&& 1 & 2 & 3 & 4 & 5 & 6 &7& & 1 & 2 & 3 & 4 & 5 & 6 &7& & 1 & 2 & 3 & 4 & 5 & 6&7 \\\hline
1&  1&  2&  \cellcolor{gray}2& 1& \cellcolor{gray}2& 3& \cellcolor{gray}3& 1&& \cellcolor{gray}1& 1& \cellcolor{gray}3& 2& \cellcolor{gray}2& 2& 1& & \cellcolor{gray}2& \cellcolor{gray}3& 2& 1& 2& \cellcolor{gray}3&2
\\
1&  1&  1&  \cellcolor{gray}2& 3& \cellcolor{gray}2& 1& \cellcolor{gray}3& 3&& \cellcolor{gray}1& 3& \cellcolor{gray}3& 1& \cellcolor{gray}2& 3& 3& & \cellcolor{gray}2& \cellcolor{gray}3& 3& 2& 1& \cellcolor{gray}3&4
\\\hline
 1&  2&  1&  3& \cellcolor{gray}1& \cellcolor{gray}3& 2& \cellcolor{gray}3& 3&& \cellcolor{gray}1& 3& 2& \cellcolor{gray}3& \cellcolor{gray}2& 1& 1& & \cellcolor{gray}2& \cellcolor{gray}2& 1& 3& 1& \cellcolor{gray}3&3
\\
1&  2&  2&  1& \cellcolor{gray}1& \cellcolor{gray}3& 1& \cellcolor{gray}3& 4&& \cellcolor{gray}1& 1& 1& \cellcolor{gray}3& \cellcolor{gray}2& 3& 4& & \cellcolor{gray}2& \cellcolor{gray}2& 3& 1& 3& \cellcolor{gray}3&4
\\\hline
1&  3&  1&  \cellcolor{gray}3& \cellcolor{gray}3& 1& 2& \cellcolor{gray}3& 5&& 2& \cellcolor{gray}3& 2& 3& \cellcolor{gray}2& \cellcolor{gray}3& 2& & \cellcolor{gray}1& 3& \cellcolor{gray}3& 1& \cellcolor{gray}2& 3&3
\\
1&  3&  2&  \cellcolor{gray}3& \cellcolor{gray}3& 2& 1& \cellcolor{gray}3& 3&& 1& \cellcolor{gray}3& 3& 2& \cellcolor{gray}2& \cellcolor{gray}3& 5& & \cellcolor{gray}1& 1& \cellcolor{gray}3& 3& \cellcolor{gray}2& 1&5
\\\hline
1&  4&  \cellcolor{gray}1&  2& \cellcolor{gray}3& 1& \cellcolor{gray}1& 3& 4&& 2& \cellcolor{gray}2& \cellcolor{gray}2& 3& 2& \cellcolor{gray}3& 2& & \cellcolor{gray}1& \cellcolor{gray}1& 2& 1& \cellcolor{gray}2& 1&4
\\
1&  4&  \cellcolor{gray}1&  1& \cellcolor{gray}3& 2& \cellcolor{gray}1& 2& 5&& 1& \cellcolor{gray}2& \cellcolor{gray}2& 1& 3& \cellcolor{gray}3& 4& & \cellcolor{gray}1& \cellcolor{gray}1& 1& 3& \cellcolor{gray}2& 3&1
\\\hline
1&  5&  \cellcolor{gray}1&  2& 1& 3& \cellcolor{gray}3& 2& \cellcolor{gray}3&& \cellcolor{gray}1& 3& 2& \cellcolor{gray}3& \cellcolor{gray}2& 2& 4& & \cellcolor{gray}1& \cellcolor{gray}2& 3& \cellcolor{gray}1& 2& 1&4
\\
1&  5&  \cellcolor{gray}1&  1& 2& 2& \cellcolor{gray}3& 3& \cellcolor{gray}3&& \cellcolor{gray}1& 2& 1& \cellcolor{gray}3& \cellcolor{gray}2& 3& 3& & \cellcolor{gray}1& \cellcolor{gray}2& 2& \cellcolor{gray}1& 1& 2&2
\\\hline
1& 6& \cellcolor{gray}1& \cellcolor{gray}3&2& 3& \cellcolor{gray}3& 1& 2&& 2& 3& 1& 2& \cellcolor{gray}3& \cellcolor{gray}3& \cellcolor{gray}3& & \cellcolor{gray}2& \cellcolor{gray}1& 3& 2& 1& \cellcolor{gray}3&2
\\
1& 6& \cellcolor{gray}1& \cellcolor{gray}3&1& 1& \cellcolor{gray}3& 2& 4&& 1& 2& 3& 3& \cellcolor{gray}3& \cellcolor{gray}3& \cellcolor{gray}3& & \cellcolor{gray}2& \cellcolor{gray}1& 1& 3& 2& \cellcolor{gray}3&1
\\\hline
1& 7& \cellcolor{gray}1& 2& \cellcolor{gray}3& 2& \cellcolor{gray}2& 3& 3& & \cellcolor{gray}2& 1& 2& 2& \cellcolor{gray}1& \cellcolor{gray}3& 2& & 2& 2& 2& \cellcolor{gray}3& 1& \cellcolor{gray}3&\cellcolor{gray}3
\\
1& 7& \cellcolor{gray}1& 3& \cellcolor{gray}3& 1& \cellcolor{gray}2& 1& 5& & \cellcolor{gray}2& 3& 3& 1& \cellcolor{gray}1& \cellcolor{gray}3& 3& & 1& 1& 3& \cellcolor{gray}3& 2& \cellcolor{gray}3&\cellcolor{gray}3
\\\hline
1& 8& \cellcolor{gray}2& 2& \cellcolor{gray}3& 3& 3& \cellcolor{gray}3& 1& & 1& \cellcolor{gray}2& \cellcolor{gray}1& 2& 2& \cellcolor{gray}3& 2& & 2& 1& \cellcolor{gray}3& \cellcolor{gray}2& 1& 3&\cellcolor{gray}5
\\
1& 8& \cellcolor{gray}2& 3& \cellcolor{gray}3& 2& 2& \cellcolor{gray}3& 5& & 2& \cellcolor{gray}2& \cellcolor{gray}1& 1& 1& \cellcolor{gray}3& 3& & 1& 2& \cellcolor{gray}3& \cellcolor{gray}2& 2& 2&\cellcolor{gray}5
\\\hline
1& 9& \cellcolor{gray}1& 2& 1& 2& \cellcolor{gray}3& \cellcolor{gray}1& 1& & \cellcolor{gray}2& 1& \cellcolor{gray}3& 2& 2& \cellcolor{gray}3& 4& & 2& 3& \cellcolor{gray}3& \cellcolor{gray}3& 1& \cellcolor{gray}3&1
\\
1& 9& \cellcolor{gray}1& 1& 3& 1& \cellcolor{gray}3& \cellcolor{gray}1& 2& & \cellcolor{gray}2& 2& \cellcolor{gray}3& 1& 3& \cellcolor{gray}3& 5& & 1& 1& \cellcolor{gray}3& \cellcolor{gray}3& 2& \cellcolor{gray}3&5
\\\hline
1& 10& \cellcolor{gray}1& 1& \cellcolor{gray}3& 2& \cellcolor{gray}2& 1& 5& & \cellcolor{gray}1& 1& \cellcolor{gray}3& \cellcolor{gray}3& 3& 1& 5& & \cellcolor{gray}1& \cellcolor{gray}1& 3& 1& \cellcolor{gray}1& 2&1
\\
1& 10& \cellcolor{gray}1& 3& \cellcolor{gray}3& 1& \cellcolor{gray}2& 3& 3& & \cellcolor{gray}1& 2& \cellcolor{gray}3& \cellcolor{gray}3& 2& 3& 4& & \cellcolor{gray}1& \cellcolor{gray}1& 1& 2& \cellcolor{gray}1& 1&4
\\\hline
1& 11& 1& \cellcolor{gray}1& 3& 2& \cellcolor{gray}1& \cellcolor{gray}3& 4& & \cellcolor{gray}1& \cellcolor{gray}2& 2& 2& \cellcolor{gray}1& 2& 3& & 1& 3& \cellcolor{gray}3& \cellcolor{gray}3& 1& 2&\cellcolor{gray}5
\\
1& 11& 2& \cellcolor{gray}1& 2& 3& \cellcolor{gray}1& \cellcolor{gray}3& 3& & \cellcolor{gray}1& \cellcolor{gray}2& 3& 1& \cellcolor{gray}1& 1& 4& & 2& 2& \cellcolor{gray}3& \cellcolor{gray}3& 2& 3&\cellcolor{gray}5
\\\hline
1& 12& 1& 2& \cellcolor{gray}3& \cellcolor{gray}1& 2& \cellcolor{gray}3& 5& & \cellcolor{gray}2& \cellcolor{gray}2& 2& 1& 2& \cellcolor{gray}3& 4& & \cellcolor{gray}2& \cellcolor{gray}1& 3& 1& 3& \cellcolor{gray}3&3
\\
1& 12& 2& 3& \cellcolor{gray}3& \cellcolor{gray}1& 1& \cellcolor{gray}3& 1& & \cellcolor{gray}2& \cellcolor{gray}2& 1& 3& 1& \cellcolor{gray}3& 1& & \cellcolor{gray}2& \cellcolor{gray}1& 1& 2& 2& \cellcolor{gray}3&2
\\\hline
1& 13& \cellcolor{gray}1& \cellcolor{gray}3& 3& \cellcolor{gray}1& 3& 1& 3& & \cellcolor{gray}1& \cellcolor{gray}1& \cellcolor{gray}1& 1& 2& 2& 1& & \cellcolor{gray}2& 3& \cellcolor{gray}3& 1& 2& \cellcolor{gray}3&4
\\
1& 13& \cellcolor{gray}1& \cellcolor{gray}3& 2& \cellcolor{gray}1& 2& 3& 2& & \cellcolor{gray}1& \cellcolor{gray}1& \cellcolor{gray}1& 2& 1& 1& 4& & \cellcolor{gray}2& 1& \cellcolor{gray}3& 2& 3& \cellcolor{gray}3&5
\\\hline
1& 14& \cellcolor{gray}1& \cellcolor{gray}2& 2& 2& 1& \cellcolor{gray}3& 4& & 1& 1& 3& \cellcolor{gray}1& \cellcolor{gray}1& \cellcolor{gray}3& 3& & \cellcolor{gray}2& 3& \cellcolor{gray}3& 2& 3& \cellcolor{gray}3&2
\\
1& 14& \cellcolor{gray}1& \cellcolor{gray}2& 1& 1& 2& \cellcolor{gray}3& 3& & 2& 3& 1& \cellcolor{gray}1& \cellcolor{gray}1& \cellcolor{gray}3& 1& & \cellcolor{gray}2& 1& \cellcolor{gray}3& 3& 2& \cellcolor{gray}3&5
\\\hline
2& 1& \cellcolor{gray}1& \cellcolor{gray}2& 1& 2& 3& \cellcolor{gray}1& 2& & \cellcolor{gray}1& \cellcolor{gray}2& 1& 3& \cellcolor{gray}3& 1& 2& & 1& 2& 2& \cellcolor{gray}3& 2& \cellcolor{gray}3&\cellcolor{gray}4
\\
2& 1& \cellcolor{gray}1& \cellcolor{gray}2& 3& 3& 2& \cellcolor{gray}1& 1& & \cellcolor{gray}1& \cellcolor{gray}2& 3& 2& \cellcolor{gray}3& 2& 1& & 2& 1& 1& \cellcolor{gray}3& 3& \cellcolor{gray}3&\cellcolor{gray}4
\\\hline
2& 2& \cellcolor{gray}1& 2& \cellcolor{gray}3& \cellcolor{gray}3& 1& 2& 2& & \cellcolor{gray}2& \cellcolor{gray}3& 2& 1& 1& \cellcolor{gray}3& 4& & \cellcolor{gray}1& 3& \cellcolor{gray}2& 3& \cellcolor{gray}2& 1&1
\\
2& 2& \cellcolor{gray}1& 1& \cellcolor{gray}3& \cellcolor{gray}3& 2& 3& 5& & \cellcolor{gray}2& \cellcolor{gray}3& 3& 2& 2& \cellcolor{gray}3& 3& & \cellcolor{gray}1& 1& \cellcolor{gray}2& 2& \cellcolor{gray}2& 2&2
\\\hline
2& 3& \cellcolor{gray}1& 1& \cellcolor{gray}3& \cellcolor{gray}3& 3& 3& 1& & \cellcolor{gray}1& 3& 1& \cellcolor{gray}2& \cellcolor{gray}1& 2& 3& & 1& 1& 3& \cellcolor{gray}2& 3& \cellcolor{gray}3&\cellcolor{gray}4
\\
2& 3& \cellcolor{gray}1& 3& \cellcolor{gray}3& \cellcolor{gray}3& 1& 1& 5& & \cellcolor{gray}1& 2& 3& \cellcolor{gray}2& \cellcolor{gray}1& 3& 4& & 2& 2& 1& \cellcolor{gray}2& 2& \cellcolor{gray}3&\cellcolor{gray}4
\\\hline
2& 4& 1& \cellcolor{gray}1& 2& \cellcolor{gray}1& 3& \cellcolor{gray}3& 4& & \cellcolor{gray}2& 1& 3& \cellcolor{gray}3& 1& \cellcolor{gray}3& 2& & \cellcolor{gray}2& 2& 3& 1& \cellcolor{gray}1& \cellcolor{gray}3&5
\\
2& 4& 2& \cellcolor{gray}1& 3& \cellcolor{gray}1& 2& \cellcolor{gray}3& 5& & \cellcolor{gray}2& 3& 1& \cellcolor{gray}3& 2& \cellcolor{gray}3& 4& & \cellcolor{gray}2& 1& 2& 3& \cellcolor{gray}1& \cellcolor{gray}3&4
\\\hline
2& 5& \cellcolor{gray}1& \cellcolor{gray}2& 3& \cellcolor{gray}1& 3& 1& 1& & \cellcolor{gray}2& 1& \cellcolor{gray}1& 1& 3& \cellcolor{gray}3& 3& & \cellcolor{gray}1& \cellcolor{gray}1& 1& 3& \cellcolor{gray}3& 1&1
\\
2& 5& \cellcolor{gray}1& \cellcolor{gray}2& 1& \cellcolor{gray}1& 1& 2& 3& & \cellcolor{gray}2& 3& \cellcolor{gray}1& 2& 1& \cellcolor{gray}3& 1& & \cellcolor{gray}1& \cellcolor{gray}1& 3& 1& \cellcolor{gray}3& 2&5
\\\hline
2& 6& \cellcolor{gray}1& \cellcolor{gray}1& 2& 3& \cellcolor{gray}3& 3& 2& & \cellcolor{gray}2& 2& 2& \cellcolor{gray}2& 2& \cellcolor{gray}3& 1& & \cellcolor{gray}1& 3& \cellcolor{gray}3& 1& \cellcolor{gray}3& 2&3
\\
2& 6& \cellcolor{gray}1& \cellcolor{gray}1& 3& 2& \cellcolor{gray}3& 2& 1& & \cellcolor{gray}2& 3& 3& \cellcolor{gray}2& 1& \cellcolor{gray}3& 2& & \cellcolor{gray}1& 1& \cellcolor{gray}3& 2& \cellcolor{gray}3& 3&1
\\\hline
2& 7& 1& \cellcolor{gray}1& \cellcolor{gray}3& 3& 1& \cellcolor{gray}3& 5& & \cellcolor{gray}2& \cellcolor{gray}2& 1& 3& 1& \cellcolor{gray}3& 3& & \cellcolor{gray}1& 2& 2& \cellcolor{gray}1& \cellcolor{gray}2& 3&3
\\
2& 7& 2& \cellcolor{gray}1& \cellcolor{gray}3& 2& 2& \cellcolor{gray}3& 2& & \cellcolor{gray}2& \cellcolor{gray}2& 3& 2& 3& \cellcolor{gray}3& 1& & \cellcolor{gray}1& 3& 1& \cellcolor{gray}1& \cellcolor{gray}2& 2&2
\\\hline
2& 8& \cellcolor{gray}2& 1& \cellcolor{gray}3& 3& 1& \cellcolor{gray}3& 2& & \cellcolor{gray}1& \cellcolor{gray}1& 2& \cellcolor{gray}1& 1& 3& 1& & 2& 1& 2& \cellcolor{gray}3& 2& \cellcolor{gray}3&\cellcolor{gray}2
\\
2& 8& \cellcolor{gray}2& 2& \cellcolor{gray}3& 2& 2& \cellcolor{gray}3& 3& & \cellcolor{gray}1& \cellcolor{gray}1& 3& \cellcolor{gray}1& 2& 1& 3& & 1& 2& 1& \cellcolor{gray}3& 3& \cellcolor{gray}3&\cellcolor{gray}2
\\\hline
2& 9& \cellcolor{gray}1& 1& 1& \cellcolor{gray}1& \cellcolor{gray}2& 3& 3& & \cellcolor{gray}2& 2& 1& 2& \cellcolor{gray}1& \cellcolor{gray}3& 2& & \cellcolor{gray}1& \cellcolor{gray}2& \cellcolor{gray}3& 2& 2& 2&2
\\
2& 9& \cellcolor{gray}1& 2& 3& \cellcolor{gray}1& \cellcolor{gray}2& 2& 2& & \cellcolor{gray}2& 1& 2& 3& \cellcolor{gray}1& \cellcolor{gray}3& 1& & \cellcolor{gray}1& \cellcolor{gray}2& \cellcolor{gray}3& 1& 1& 3&5
\\\hline
2& 10& 2& 1& 1& \cellcolor{gray}1& \cellcolor{gray}1& \cellcolor{gray}3& 4& & \cellcolor{gray}1& \cellcolor{gray}1& 2& 1& \cellcolor{gray}2& 3& 2& & \cellcolor{gray}1& 1& \cellcolor{gray}3& 3& \cellcolor{gray}3& 2&3
\\
2& 10& 1& 2& 2& \cellcolor{gray}1& \cellcolor{gray}1& \cellcolor{gray}3& 1& & \cellcolor{gray}1& \cellcolor{gray}1& 1& 3& \cellcolor{gray}2& 2& 4& & \cellcolor{gray}1& 3& \cellcolor{gray}3& 1& \cellcolor{gray}3& 1&5
\\\hline
2& 11& \cellcolor{gray}1& 3& 1& \cellcolor{gray}3& \cellcolor{gray}1& 2& 2& & \cellcolor{gray}2& 1& 1& \cellcolor{gray}2& 2& \cellcolor{gray}3& 2& & \cellcolor{gray}1& \cellcolor{gray}2& 1& 1& \cellcolor{gray}3& 2&4
\\
2& 11& \cellcolor{gray}1& 1& 3& \cellcolor{gray}3& \cellcolor{gray}1& 1& 3& & \cellcolor{gray}2& 2& 2& \cellcolor{gray}2& 1& \cellcolor{gray}3& 4& & \cellcolor{gray}1& \cellcolor{gray}2& 3& 2& \cellcolor{gray}3& 1&2
\\\hline
2& 12& \cellcolor{gray}2& 1& 1& 3& \cellcolor{gray}3& \cellcolor{gray}3& 1& & 1& 3& 3& \cellcolor{gray}2& 1& \cellcolor{gray}3& \cellcolor{gray}2& & \cellcolor{gray}2& 3& \cellcolor{gray}3& 1& 1& \cellcolor{gray}3&5
\\
2& 12& \cellcolor{gray}2& 2& 3& 1& \cellcolor{gray}3& \cellcolor{gray}3& 4& & 2& 1& 2& \cellcolor{gray}2& 3& \cellcolor{gray}3& \cellcolor{gray}2& & \cellcolor{gray}2& 2& \cellcolor{gray}3& 2& 2& \cellcolor{gray}3&3
\\\hline
2& 13& 1& 3& \cellcolor{gray}3& \cellcolor{gray}2& 3& \cellcolor{gray}3& 3& & \cellcolor{gray}1& \cellcolor{gray}1& 3& \cellcolor{gray}3& 1& 3& 2& & 1& 3& 2& \cellcolor{gray}1& 3& \cellcolor{gray}3&\cellcolor{gray}1
\\
2& 13& 2& 2& \cellcolor{gray}3& \cellcolor{gray}2& 2& \cellcolor{gray}3& 1& & \cellcolor{gray}1& \cellcolor{gray}1& 2& \cellcolor{gray}3& 2& 1& 4& & 2& 2& 3& \cellcolor{gray}1& 1& \cellcolor{gray}3&\cellcolor{gray}1
\\\hline
2& 14& 1& \cellcolor{gray}2& 1& 3& \cellcolor{gray}3& \cellcolor{gray}3& 1& & \cellcolor{gray}2& 2& 2& \cellcolor{gray}3& 2& \cellcolor{gray}3& 2& & \cellcolor{gray}1& \cellcolor{gray}2& \cellcolor{gray}1& 3& 1& 3&4
\\
2& 14& 2& \cellcolor{gray}2& 2& 1& \cellcolor{gray}3& \cellcolor{gray}3& 3& & \cellcolor{gray}2& 1& 1& \cellcolor{gray}3& 3& \cellcolor{gray}3& 4& & \cellcolor{gray}1& \cellcolor{gray}2& \cellcolor{gray}1& 2& 3& 1&3
\\\hline
3& 1& \cellcolor{gray}1& \cellcolor{gray}2& 2& \cellcolor{gray}2& 2& 2& 4& & \cellcolor{gray}1& 2& 2& \cellcolor{gray}1& \cellcolor{gray}1& 2& 2& & 1& 2& 1& \cellcolor{gray}1& 3& \cellcolor{gray}3&\cellcolor{gray}1
\\
3& 1& \cellcolor{gray}1& \cellcolor{gray}2& 3& \cellcolor{gray}2& 3& 3& 2& & \cellcolor{gray}1& 3& 1& \cellcolor{gray}1& \cellcolor{gray}1& 3& 1& & 2& 3& 2& \cellcolor{gray}1& 1& \cellcolor{gray}3&\cellcolor{gray}1
\\\hline
3& 2& \cellcolor{gray}1& \cellcolor{gray}2& \cellcolor{gray}3& 3& 2& 1& 4& & \cellcolor{gray}2& 2& \cellcolor{gray}3& 1& 2& \cellcolor{gray}3& 1& & \cellcolor{gray}2& 3& \cellcolor{gray}3& 2& 2& \cellcolor{gray}3&1
\\
3& 2& \cellcolor{gray}1& \cellcolor{gray}2& \cellcolor{gray}3& 2& 3& 3& 5& & \cellcolor{gray}2& 1& \cellcolor{gray}3& 2& 1& \cellcolor{gray}3& 5& & \cellcolor{gray}2& 2& \cellcolor{gray}3& 3& 1& \cellcolor{gray}3&4
\\\hline
3& 3& \cellcolor{gray}1& 1& 3& \cellcolor{gray}2& \cellcolor{gray}3& 2& 2& & \cellcolor{gray}2& 3& \cellcolor{gray}1& 1& 3& \cellcolor{gray}3& 2& & \cellcolor{gray}1& \cellcolor{gray}3& 3& 2& \cellcolor{gray}1& 2&1
\\
3& 3& \cellcolor{gray}1& 3& 1& \cellcolor{gray}2& \cellcolor{gray}3& 1& 4& & \cellcolor{gray}2& 2& \cellcolor{gray}1& 2& 1& \cellcolor{gray}3& 3& & \cellcolor{gray}1& \cellcolor{gray}3& 2& 3& \cellcolor{gray}1& 3&2
\\\hline
3& 4& \cellcolor{gray}2& 3& 2& 1& \cellcolor{gray}1& \cellcolor{gray}3& 1& & \cellcolor{gray}1& \cellcolor{gray}3& \cellcolor{gray}3& 3& 1& 2& 1& & 1& 2& 2& \cellcolor{gray}2& 1& \cellcolor{gray}3&\cellcolor{gray}1
\\
3& 4& \cellcolor{gray}2& 2& 1& 2& \cellcolor{gray}1& \cellcolor{gray}3& 2& & \cellcolor{gray}1& \cellcolor{gray}3& \cellcolor{gray}3& 2& 2& 1& 5& & 2& 1& 1& \cellcolor{gray}2& 3& \cellcolor{gray}3&\cellcolor{gray}1
\\\hline
3& 5& \cellcolor{gray}1& 1& \cellcolor{gray}1& \cellcolor{gray}1& 3& 2& 4& & 1& 3& 2& \cellcolor{gray}2& \cellcolor{gray}2& \cellcolor{gray}3& 3& & \cellcolor{gray}1& \cellcolor{gray}3& 1& \cellcolor{gray}3& 1& 3&3
\\
3& 5& \cellcolor{gray}1& 3& \cellcolor{gray}1& \cellcolor{gray}1& 1& 3& 1& & 2& 2& 3& \cellcolor{gray}2& \cellcolor{gray}2& \cellcolor{gray}3& 5& & \cellcolor{gray}1& \cellcolor{gray}3& 2& \cellcolor{gray}3& 3& 2&1
\\\hline
3& 6& \cellcolor{gray}1& 2& \cellcolor{gray}3& \cellcolor{gray}2& 1& 1& 5& & 2& \cellcolor{gray}1& \cellcolor{gray}3& 1& 2& \cellcolor{gray}3& 5& & \cellcolor{gray}2& 1& 2& 2& \cellcolor{gray}3& \cellcolor{gray}3&3
\\
3& 6& \cellcolor{gray}1& 1& \cellcolor{gray}3& \cellcolor{gray}2& 2& 2& 2& & 1& \cellcolor{gray}1& \cellcolor{gray}3& 3& 3& \cellcolor{gray}3& 2& & \cellcolor{gray}2& 2& 1& 3& \cellcolor{gray}3& \cellcolor{gray}3&2
\\\hline
3& 7& \cellcolor{gray}1& 2& \cellcolor{gray}3& \cellcolor{gray}3& 1& 3& 5& & \cellcolor{gray}2& 3& 2& 2& \cellcolor{gray}1& \cellcolor{gray}3& 3& & 1& 3& 2& \cellcolor{gray}2& 3& \cellcolor{gray}3&\cellcolor{gray}3
\\
3& 7& \cellcolor{gray}1& 3& \cellcolor{gray}3& \cellcolor{gray}3& 3& 1& 2& & \cellcolor{gray}2& 2& 1& 3& \cellcolor{gray}1& \cellcolor{gray}3& 4& & 2& 1& 3& \cellcolor{gray}2& 2& \cellcolor{gray}3&\cellcolor{gray}3
\\\hline
3& 8& \cellcolor{gray}1& 1& 1& 3& \cellcolor{gray}3& \cellcolor{gray}2& 3& & \cellcolor{gray}1& 1& 3& \cellcolor{gray}3& \cellcolor{gray}2& 2& 4& & \cellcolor{gray}1& \cellcolor{gray}1& 1& 2& \cellcolor{gray}1& 3&4
\\
3& 8& \cellcolor{gray}1& 3& 2& 2& \cellcolor{gray}3& \cellcolor{gray}2& 1& & \cellcolor{gray}1& 2& 3& \cellcolor{gray}3& \cellcolor{gray}3& 3& 5& & \cellcolor{gray}1& \cellcolor{gray}1& 2& 3& \cellcolor{gray}1& 1&3
\\\hline
3& 9& 1& 3& \cellcolor{gray}3& 2& \cellcolor{gray}2& \cellcolor{gray}3& 4& & 1& 3& 1& \cellcolor{gray}1& 3& \cellcolor{gray}3& \cellcolor{gray}1& & \cellcolor{gray}1& \cellcolor{gray}1& \cellcolor{gray}1& 2& 1& 2&3
\\
3& 9& 2& 2& \cellcolor{gray}3& 1& \cellcolor{gray}2& \cellcolor{gray}3& 5& & 2& 1& 3& \cellcolor{gray}1& 1& \cellcolor{gray}3& \cellcolor{gray}1& & \cellcolor{gray}1& \cellcolor{gray}1& \cellcolor{gray}1& 1& 3& 3&2
\\\hline
3& 10& \cellcolor{gray}1& \cellcolor{gray}3& 1& \cellcolor{gray}1& 3& 3& 4& & \cellcolor{gray}1& \cellcolor{gray}3& \cellcolor{gray}3& 2& 3& 1& 1& & \cellcolor{gray}2& 3& 1& 1& \cellcolor{gray}1& \cellcolor{gray}3&3
\\
3& 10& \cellcolor{gray}1& \cellcolor{gray}3& 2& \cellcolor{gray}1& 1& 2& 3& & \cellcolor{gray}1& \cellcolor{gray}3& \cellcolor{gray}3& 1& 1& 2& 5& & \cellcolor{gray}2& 2& 2& 2& \cellcolor{gray}1& \cellcolor{gray}3&1
\\\hline
3& 11& \cellcolor{gray}1& 1& 2& \cellcolor{gray}3& \cellcolor{gray}1& 2& 4& & \cellcolor{gray}1& \cellcolor{gray}3& \cellcolor{gray}3& 1& 3& 2& 4& & \cellcolor{gray}2& \cellcolor{gray}3& 2& 1& 3& \cellcolor{gray}3&4
\\
3& 11& \cellcolor{gray}1& 3& 1& \cellcolor{gray}3& \cellcolor{gray}1& 1& 2& & \cellcolor{gray}1& \cellcolor{gray}3& \cellcolor{gray}3& 3& 2& 1& 1& & \cellcolor{gray}2& \cellcolor{gray}3& 3& 3& 1& \cellcolor{gray}3&2
\\\hline
3& 12& \cellcolor{gray}1& \cellcolor{gray}1& 3& \cellcolor{gray}1& 1& 3& 4& & 2& \cellcolor{gray}2& \cellcolor{gray}3& 2& 3& \cellcolor{gray}3& 3& & \cellcolor{gray}1& 3& \cellcolor{gray}3& 2& \cellcolor{gray}2& 3&5
\\
3& 12& \cellcolor{gray}1& \cellcolor{gray}1& 2& \cellcolor{gray}1& 3& 1& 3& & 1& \cellcolor{gray}2& \cellcolor{gray}3& 3& 2& \cellcolor{gray}3& 5& & \cellcolor{gray}1& 2& \cellcolor{gray}3& 3& \cellcolor{gray}2& 1&4
\\\hline
3& 13& \cellcolor{gray}1& \cellcolor{gray}2& \cellcolor{gray}3& 3& 2& 1& 3& & \cellcolor{gray}1& 2& \cellcolor{gray}2& \cellcolor{gray}1& 1& 1& 3& & \cellcolor{gray}2& 3& \cellcolor{gray}3& 2& 2& \cellcolor{gray}3&5
\\
3& 13& \cellcolor{gray}1& \cellcolor{gray}2& \cellcolor{gray}3& 1& 3& 2& 5& & \cellcolor{gray}1& 1& \cellcolor{gray}2& \cellcolor{gray}1& 2& 3& 2& & \cellcolor{gray}2& 1& \cellcolor{gray}3& 3& 3& \cellcolor{gray}3&1
\\\hline
3& 14& \cellcolor{gray}1& 2& 2& \cellcolor{gray}3& \cellcolor{gray}1& 1& 2& & \cellcolor{gray}1& 2& 1& \cellcolor{gray}3& \cellcolor{gray}3& 2& 2& & 1& 3& 3& \cellcolor{gray}1& 2& \cellcolor{gray}3&\cellcolor{gray}2
\\
 3& 14& \cellcolor{gray}1& 3& 3& \cellcolor{gray}3& \cellcolor{gray}1& 2& 1& & \cellcolor{gray}1& 1& 2& \cellcolor{gray}3& \cellcolor{gray}3& 3& 3& & 2& 2& 1& \cellcolor{gray}1& 3& \cellcolor{gray}3&\cellcolor{gray}2
\\
\hline
    \end{longtable}
    \end{center}
  
\end{document}